\pdfoutput=1
\documentclass[11pt,twoside,a4paper,cmspaper,final,collab]{cms-tdr}
\def\svnVersion{e289ab3}\def\svnDate{2021/11/09}\def\cmsCernNoTag{CERN-EP-2021-120}\def\cmsCernDate{\today}\def\cmsMessage{Published in the European Physical Journal C as \href{http://dx.doi.org/10.1140/epjc/s10052-021-09721-5}{\doi{10.1140/epjc/s10052-021-09721-5}.}}
\begin{document}\cmsNoteHeader{SUS-20-002}

\newlength\cmsFigWidth
\ifthenelse{\boolean{cms@external}}{\setlength\cmsFigWidth{0.98\columnwidth}}{\setlength\cmsFigWidth{0.49\textwidth}}
\newlength\cmsFigWidthTwo
\ifthenelse{\boolean{cms@external}}{\setlength\cmsFigWidthTwo{0.66\columnwidth}}{\setlength\cmsFigWidthTwo{0.32\textwidth}}
\newlength\cmsTabSkip\setlength{\cmsTabSkip}{1ex}
\newcommand{\amcatnlo}{\MGvATNLO}

\newcommand{\mtll}{\ensuremath{\mTii(\ell\ell)}\xspace}
\newcommand{\mtlblb}{\ensuremath{\mTii(\PQb\ell \PQb\ell )}\xspace}

\newcommand{\mtop}{\ensuremath{m_\PQt}\xspace}
\newcommand{\tW}{\ensuremath{\PQt\PW}\xspace}
\newcommand{\ttW}{\ensuremath{\PQt\PAQt\PW}\xspace}
\newcommand{\ttZ}{\ensuremath{\PQt\PAQt\PZ}\xspace}
\newcommand{\tqZ}{\ensuremath{\PQt\PQq\PZ}\xspace}
\newcommand{\ttH}{\ensuremath{\ttbar\PH}\xspace}
\newcommand{\ttDM}{\ensuremath{\ttbar+\text{DM}}\xspace}
\newcommand{\pchi}{\PSGczDo}
\newcommand{\pstop}{\PSQtDo}
\newcommand{\mchi}{\ensuremath{{m}_{\pchi}}\xspace}
\newcommand{\mstop}{\ensuremath{{m}_{\pstop}}\xspace}
\newcommand{\deltamsn}{\ensuremath{\Delta m\left(\pstop,\pchi\right)}\xspace}
\newcommand{\deltamcor}{\ensuremath{\Delta m_\text{cor}}\xspace}

\newcommand{\mlsp}{\mchi}
\newcommand{\mll}{\ensuremath{m_{\ell\ell}\xspace}}
\newcommand{\ee}{\ensuremath{{\Pe\Pe}}\xspace}
\newcommand{\mumu}{\ensuremath{\PGm\PGm}\xspace}
\newcommand{\emu}{\ensuremath{\Pe\PGm}\xspace}
\newcommand{\wjets}{\PW{}+jets\xspace}
\newcommand{\zjets}{\PZ{}+jets\xspace}
\newcommand{\gjets}{\PGg{}+jets\xspace}
\newcommand{\tmod}{\ensuremath{t_{\text{mod}}}\xspace}
\newcommand{\mW}{\ensuremath{m_{\PW}}\xspace}
\newcommand{\ptmissSig}{\ensuremath{\mathcal{S}}\xspace}

\cmsNoteHeader{SUS-20-002}
\title{Combined searches for the production of supersymmetric top quark partners in proton-proton collisions at \texorpdfstring{$\sqrt{s} = 13\TeV$}{sqrt(s) = 13 TeV} }
\titlerunning{Combined searches for the production of supersymmetric top quark partners\ldots at 13\TeV}
\date{\today}

\abstract{A combination of searches for top squark pair production using proton-proton collision data at a center-of-mass energy of 13\TeV at the CERN LHC, corresponding to an integrated luminosity of 137\fbinv collected by the CMS experiment, is presented. Signatures with at least 2 jets and large missing transverse momentum are categorized into events with 0, 1, or 2 leptons. New results for regions of parameter space where the kinematical properties of top squark pair production and top quark pair production are very similar are presented.
Depending on the model, the combined result excludes a top squark mass up to 1325\GeV for a massless neutralino, and a neutralino mass up to 700\GeV for a top squark mass of 1150\GeV. Top squarks with masses from 145 to 295\GeV, for neutralino masses from 0 to 100\GeV, with a mass difference between the top squark and the neutralino in a window of 30\GeV around the mass of the top quark, are excluded for the first time with CMS data.
The results of theses searches are also interpreted in an alternative signal model of dark matter production via a spin-0 mediator in association with a top quark pair. Upper limits are set on the cross section for mediator particle masses of up to 420\GeV.}

\hypersetup{%
pdfauthor={CMS Collaboration},%
pdftitle={Combined searches for the production of supersymmetric top quark partners in proton-proton collisions at sqrt(s)=13 TeV},
pdfsubject={CMS},%
pdfkeywords={CMS, supersymmetry}}

\maketitle

\section{Introduction}\label{intro}

The standard model (SM) of particle physics describes subatomic phenomena with outstanding precision. However, the SM cannot address several open questions such as the hierarchy problem~\cite{Barbieri:1987fn, WITTEN1981513} and the absence of a suitable particle candidate for dark matter (DM)~\cite{Bertone:2004pz, DarkMatter2}.
Supersymmetry (SUSY)~\cite{Ramond:1971gb,Golfand:1971iw,Neveu:1971rx,Volkov:1972jx,Wess:1973kz,Wess:1974tw,Fayet:1974pd,Nilles:1983ge} is a well-known extension of the SM that can resolve both of these problems by introducing a relation between bosons and fermions. For each known particle, it assigns a new SUSY partner that differs by a half unit of spin. 
SUSY provides a natural solution to the gauge hierarchy problem provided that the SUSY partners of the top quark, gluon, and Higgs boson are not too massive.
While difficult to quantify precisely, values of the top squark mass up to the 1\TeV range are favored~\cite{Faraggi:2000pv,Barbieri:2009ev,Papucci:2011wy,Barbieri:1987fn}.
The lightest SUSY particle (LSP), which is potentially massive, may be a viable DM candidate if it is stable and electrically neutral.

This paper presents the combination of previously published searches~\cite{Sirunyan:2021mrs,Sirunyan:2019glc,Sirunyan:2020tyy} for the pair production of SUSY top quark partners in final states without leptons, with one, or with two charged leptons, in events from proton-proton ($\Pp\Pp$)~collisions at a center-of-mass energy ($\sqrt{s}$) of 13\TeV at the CERN LHC, corresponding to an integrated luminosity of 137\fbinv,  and referred to here as inclusive analyses. It also includes a new analysis targeting a parameter space where the mass difference between the top squark and the neutralino is close to the top quark mass, whose results are combined with the previously published studies. All analyses are performed with the data set collected in 2016--2018 (Run 2) by CMS.

\begin{figure*}[htb!]
  \centering
    \includegraphics[width=.32\linewidth]{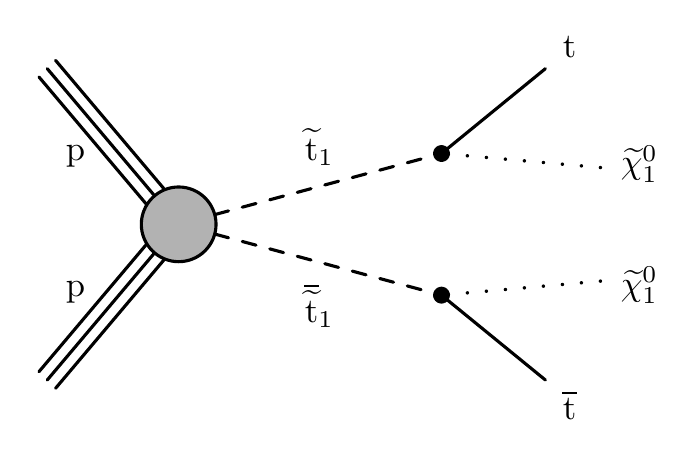}
		\includegraphics[width=.32\linewidth]{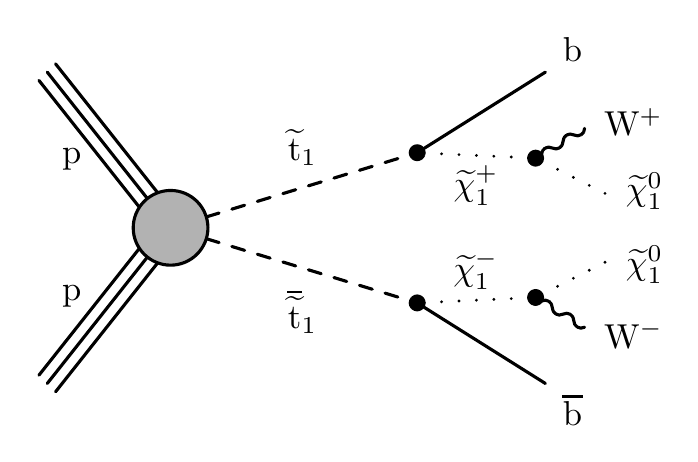}
		\includegraphics[width=.32\linewidth]{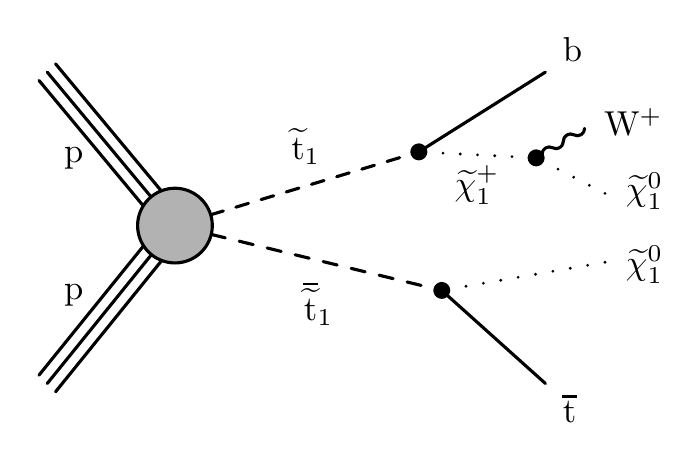}
    \caption{Diagrams of top squark pair production with further decay of each top squark into a top quark and a neutralino (left), of each top squark into a chargino and a neutralino, with the chargino decaying then into a bottom quark and a \PW boson (center), and with a combination of the two top squark decay scenarios (right).}
    \label{fig:diagrams}
\end{figure*}

The inclusive searches are interpreted in terms of top squark pair production with two different subsequent decays, as described in the simplified model context~\cite{Alwall:2008ag,Alwall:2008va,Alves:2011wf}.
Two decay chains are considered, both of which lead to a signature with a neutralino (\PSGczDo), which is the LSP, a \PW boson and a bottom quark. These are the direct decay of the top squark to a top quark and a neutralino, and the decay of the top squark to a chargino (\PSGcpmDo) and a bottom quark where the chargino decays to a \PW boson and a neutralino. Three simplified models are used for interpretation. In the first model, both top squarks decay according to the first decay chain; in the second model, both decay according to the second decay chain; in the third model, these two decays occur with equal probability. The mass of the chargino in the second model is chosen to be an arithmetic average of the top squark mass (\mstop) and the LSP mass (\mchi), while in the third model the mass splitting between the neutralino and chargino is assumed to be 5\GeV.
Typical diagrams are shown in Fig.~\ref{fig:diagrams}.
In previous analyses by the CMS collaboration top squark masses up to 1310\GeV have been excluded \cite{Sirunyan:2021mrs,Sirunyan:2019glc,Sirunyan:2020tyy,Sirunyan:2017xse,Sirunyan:2017leh,Chatrchyan:2013xna,Khachatryan:2016pup,Khachatryan:2016pxa,Sirunyan:2016jpr,Sirunyan:2017wif,Sirunyan:2017pjw}. Limits on the production of top squark pairs with masses up to 1260\GeV have been set by the ATLAS Collaboration~\cite{Aad:2021nfd,Aad:2015pfx,Aad:2014kra,Aad:2014qaa,Aad:2021lwz,Aad:2021had}.

If the mass difference between the top squark and the lightest neutralino in the $\PSQtDo \to \PQt \PSGczDo$ model is close to the mass of the top quark (\mtop), the kinematic distributions of the final states of the SUSY signal are very similar to those of SM top quark pair (\ttbar) production processes. Therefore, this is a difficult region in which to search for a signal.
In this case, the signal acceptance strongly depends on $m_{\PSQtDo}$ and $m_{\PSGczDo}$. The boundaries of the corridor are taken to be $\deltamcor = 30\GeV$ and $m_{\PSQtDo}\lesssim275\GeV$, where $\deltamcor \equiv \vert \deltamsn - 175\GeV\vert$ and 175\GeV is the reference value of the top quark mass. 
The top quark corridor was not included in the parameter space addressed by the previous inclusive searches by the CMS Collaboration~\cite{Sirunyan:2021mrs,Sirunyan:2019glc,Sirunyan:2020tyy,Sirunyan:2017xse,Sirunyan:2017leh,Chatrchyan:2013xna,Khachatryan:2016pup,Khachatryan:2016pxa,Sirunyan:2016jpr,Sirunyan:2017wif,Sirunyan:2017pjw}.

In the top quark corridor region, the signal could be observed as an excess over the \ttbar background prediction \cite{Khachatryan:2016mqs}, but the sensitivity to $\mlsp\geq 20\GeV$ is limited. A dedicated search was performed with the data set collected in 2016 by CMS~\cite{Sirunyan:2019zyu}, that excluded the presence of top squark production up to $\mstop = 240\GeV$ for $\deltamcor = 0$ and up to about $\mstop = 208\GeV$ for $\deltamcor = 7.5\GeV$ at 95\% confidence level. An analysis of the top quark corridor by the ATLAS Collaboration has set exclusion limits for top squark masses between 170 and 230\GeV\cite{Aaboud:2019hwz}. 

This paper presents a new dedicated search in events with an opposite-charge lepton pair that is sensitive to the top quark corridor region. 
The sensitivity in the top quark corridor is extended by using a larger data set and a more sophisticated strategy, using a deep neural network (DNN)~\cite{Baldi:2016fzo} to exploit the  differences between the signal and the SM \ttbar production, which is the main background. 

In order to reduce the background from \ttbar events, the missing transverse momentum (\ptvecmiss) is used together with the so-called ``stransverse'' mass of the leptons (\mtll)~\cite{Christopher:1999mt2}, defined as
\begin{linenomath}
\ifthenelse{\boolean{cms@external}}
{
\begin{multline*}
\mtll = \min_{\vec{p}_{\text{T1}}^{\text{miss}} + \vec{p}_{\text{T2}}^{\text{miss}} = \ptvecmiss} \\
\left( \max \left[ \mT(\pt^{{\ell}1},\vec{p}_{\text{T1}}^{\text{miss}}), \mT(\pt^{{\ell}2},\vec{p}_{\text{T2}}^{\text{miss}}) \right] \right),
\end{multline*}
}
{
\begin{equation*}
\mtll = \min_{\vec{p}_{\text{T1}}^{\text{miss}} + \vec{p}_{\text{T2}}^{\text{miss}} = \ptvecmiss} \left( \max \left[ \mT(\pt^{{\ell}1},\vec{p}_{\text{T1}}^{\text{miss}}) , \mT(\pt^{{\ell}2},\vec{p}_{\text{T2}}^{\text{miss}}) \right] \right),
\end{equation*}
}
\end{linenomath}\label{eq:mtll}
where $\ell$ refers to an electron or a muon, \mT is the transverse mass, and $\vec{p}_\text{T1}^\text{miss}$ and $\vec{p}_\text{T2}^\text{miss}$ correspond to the estimated transverse momenta of the two invisible particles (neutrinos in the case of \ttbar events) that are presumed to determine the total \ptvecmiss of an SM event. The transverse mass is calculated for each lepton-neutrino pair, for different assumptions of the neutrino transverse momentum ($\vec{p}_{\text{T}i}^{\text{miss}}$). The computation of $\mtll$ is done using the algorithm discussed in Ref. \cite{Cheng:2008hk}. A signal region is defined applying requirements on \mtll and on \ptmiss, the magnitude of \ptvecmiss. A DNN is used to optimize the sensitivity for signal at each mass point.

\begin{figure}[htb!]
  \centering
    \includegraphics[width=\cmsFigWidthTwo]{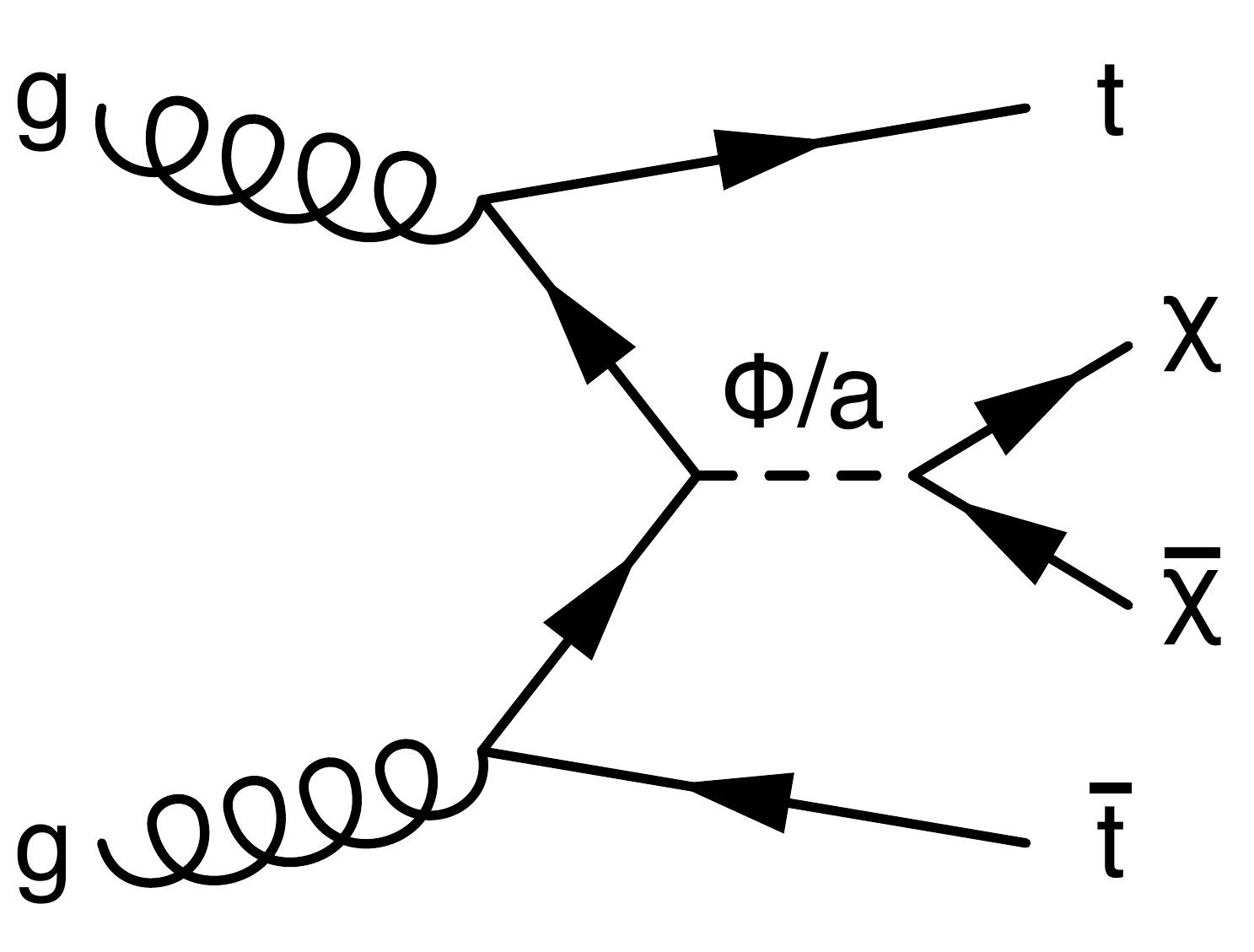}
    \caption{Feynman diagram of direct DM production through a scalar ($\phi$) or pseudoscalar (a) mediator particle, in association with a top quark pair.}
    \label{fig:TTDMdiagrams}
\end{figure}

We also consider the alternative model \ttDM shown in Fig.~\ref{fig:TTDMdiagrams}, in which a DM particle is produced in association with a pair of top quarks.
In this simplified model, a scalar ($\phi$) or pseudoscalar (a) particle mediates the interaction between SM quarks and a new Dirac fermion ($\chi$), which is the DM candidate particle~\cite{Lin:2013sca,Buckley:2014fba,Haisch:2015ioa,Arina:2016cqj,Abercrombie:2015wmb}.
Under the assumption of minimal flavor violation~\cite{DAmbrosio:2002vsn,Isidori:2012ts} the spin-0 mediators couple to quarks having mass $m_\Pq$ with SM-like Yukawa couplings proportional to $g_\Pq m_\Pq$, where the coupling strength $g_\Pq$ is taken to be 1. The coupling strength $g_{\text{DM}}$ of the mediator to the DM particles is also set to 1.
In the case of a scalar mediator, mixing with the SM Higgs boson is neglected.
Prior searches by the ATLAS and CMS Collaborations excluded scalar and pseudoscalar mediator particles with a mass of up to 290 and 300\GeV, respectively~\cite{Aaboud:2017rzf,Aad:2021nfd,Sirunyan:2018dub,Sirunyan:2019gfm,top18003}.

\section{The CMS detector}\label{cms}
The central feature of the CMS apparatus is a superconducting solenoid of 6\unit{m} internal diameter, providing a magnetic field of 3.8\unit{T}. Within the solenoid volume are a silicon pixel and strip tracker, a lead tungstate crystal electromagnetic calorimeter (ECAL), and a brass and scintillator hadron calorimeter (HCAL), each composed of a barrel and two endcap sections. Forward calorimeters extend the pseudorapidity coverage provided by the barrel and endcap detectors. Muons are detected in gas-ionization chambers embedded in the steel flux-return yoke outside the solenoid.

Events of interest are selected using a two-tiered trigger system. The first level, composed of custom hardware processors, uses information from the calorimeters and muon detectors to select events at a rate of around 100\unit{kHz} within a fixed latency of about 4\mus~\cite{Sirunyan:2020tr}. The second level, known as the high-level trigger, consists of a farm of processors running a version of the full event reconstruction software optimized for fast processing, and reduces the event rate to around 1\unit{kHz} before data storage~\cite{Khachatryan:2016bia}.

A more detailed description of the CMS detector, together with a definition of the coordinate system used and the relevant kinematic variables, can be found in Ref.~\cite{bib:cms}.

\section{Monte Carlo simulation}\label{sec:MC}
Monte Carlo (MC) simulation is used to design the searches, predict or aid the prediction of the background events from SM processes, and to provide estimations of the expected SUSY and \ttDM signal event yields.

Several models from the simplified model spectra~\cite{Alwall:2008ag, Alves:2011wf} are used to simulate the SUSY signals. The helicity states of the produced top quarks are not considered in these models, and in the simulation the top quarks are treated as unpolarized. The generation of signal samples is performed using the \amcatnlo~generator (\MADGRAPH)~\cite{Alwall:2014hca, Alwall:2007fs} (version 2.2.2 for 2016 and  version 2.4.2 for 2017 and 2018) at leading order (LO) in quantum chromodynamics (QCD) with up to two additional partons from initial-state radiation (ISR). To improve on the \MADGRAPH modeling of the multiplicity of additional jets from ISR, \MADGRAPH signal events are reweighted based on the number of ISR jets ($N_\mathrm{J}^\mathrm{ISR}$). These weights are obtained using a \ttbar \MADGRAPH MC sample, so as to make the \ttbar jet multiplicity agree with data. The reweighting factors vary between 0.92 and 0.51 for $N_\mathrm{J}^\mathrm{ISR}$ between 1 and 6, respectively.

Signal samples of the \ttDM model~\cite{Mattelaer:2015haa} are generated using \MADGRAPH~v2.4.2 at LO with at most one additional parton in the matrix element calculations. Samples for mediator masses of 50, 100, 200, 300, and 500\GeV have been generated for both the scalar and pseudoscalar models. The mass of the DM particle is set to 1\GeV while a value of 1 is chosen for the couplings.

The SM \ttbar process is simulated using the \POWHEG (v2)~\cite{bib:powheg2,Frixione:2007vw,Nason:2004rx} generator at next-to-leading order (NLO) for the dilepton analyses or the \MADGRAPH~generator at LO for the analyses of zero or one lepton events. In the top quark corridor analysis the \POWHEG generator is used, as this analysis relies on a precise estimate of the \ttbar background and its associated modeling uncertainties, which are better described in CMS by the \POWHEG generator~\cite{Khachatryan:2016mqs,Sirunyan:2018ptc}.
This sample is also used to calculate the dependence of the \ttbar acceptance on \mtop and on the factorization and renormalization scales ($\mu_\mathrm{F}$, $\mu_\mathrm{R}$, respectively). A parameter denoted as $h_\mathrm{damp}$ is used in the modeling of the parton shower matrix element \cite{CMS-PAS-TOP-16-021, Sirunyan:2019dfx}. The central value and uncertainties in $h_{\mathrm{damp}}$ are discussed in Section \ref{sec:ttmodelling}. 

The \POWHEG v1~\cite{bib:powheg3} generator is used for the single top quark and antiquark production in association with a \PW~boson (\tW) at NLO. The \MADGRAPH~v2.2.2~\cite{Alwall:2014hca} generator is used at NLO for modeling the Drell--Yan (DY) process, the production of \PW\ or \PZ\ bosons in association with \ttbar events ($\ttbar\PW$, $\ttbar\PZ$), and the $\PW\PW$, $\PW\PZ$, and $\PZ\PZ$ production processes. The production of the DY process is simulated with up to two additional partons~\cite{Frederix:2012ps}, and the FxFx scheme is used for the matching of jets from the matrix element calculations and from parton showers.
Samples of \wjets, \zjets events (with $\PZ \to {\PGn}{\PAGn}$), \gjets, and QCD multijet production are simulated with up to four extra partons in the matrix element calculations using the \MADGRAPH~(v2.2.2 in 2016 and v2.4.2 in 2017 and 2018) event generator at LO.
Double counting of the partons generated with \MADGRAPH and via the parton shower is removed using the MLM~\cite{Alwall:2007fs} matching scheme.

The NNPDF 3.0~\cite{Ball:2014uwa} parton distribution function (PDF) set is used for generating the samples corresponding to the 2016 period, while the NNPDF 3.1 NNLO~\cite{Ball:2017pdf} PDF is used for the 2017 and 2018 samples. Parton showering and hadronization are handled by \PYTHIA~v8.226 (8.230)~\cite{Sjostrand:2014zea,Skands:2014pea} using the underlying event tune CUETP8M2T4~\cite{CMS-PAS-TOP-16-021} for SM \ttbar~events for the 2016 (2017, 2018) period, the CUETP8M1~\cite{Khachatryan:2015pea} tune for all other background and signal events in the 2016 period, and the CP5 tune~\cite{Sirunyan:2019dfx} for all background and signal events of the 2017 and 2018 periods. The nominal top quark mass is 172.5\GeV in all the samples.

The \GEANTfour package~\cite{Agostinelli:2002hh} is used to simulate the CMS detector for samples of the SM processes, the \ttDM signal processes, and SUSY signal samples where $m_{\PSQtDo}-m_{\PSGczDo}$ is close to the top quark mass.
The CMS fast simulation program~\cite{Abdullin:2011zz,Giammanco:2014bza} is used to simulate the detector response for the remaining signal samples. The effect of additional interactions in the same event (referred to as pileup) is accounted for by simulating additional minimum bias interactions for each hard scattering event. The observed distribution of the number of pileup interactions, which has an average of 23 and 32 collisions per bunch crossing for the 2016 period, and for the 2017 and 2018 periods, respectively, is reproduced by the simulation.

Simulated background events are normalized according to the integrated luminosity and the theoretical cross section of each process. The latter are computed at next-to-next-to-leading order (NNLO) in QCD for DY~\cite{PhysRevD.86.094034}, approximately NNLO for \tW~\cite{Kidonakis:2015nna}, and NLO for $\PW\PW$, $\PW\PZ$, $\PZ\PZ$~\cite{bib:mcfm:diboson}, $\ttbar\PW$ and $\ttbar\PZ$~\cite{Garzelli:2012bn}. For the normalization of the simulated \ttbar sample, the full NNLO plus next-to-next-to-leading logarithmic (NNLL) accurate calculation~\cite{Czakon:2013goa}, performed with the \textsc{Top++} 2.0 program~\cite{Czakon:2011xx}, is used. The PDF uncertainties are added in quadrature to the uncertainty associated with the strong coupling constant (\alpS) to obtain a \ttbar production cross section of $832~^{+20}_{-29}\,\text{(scale)}\pm 35\,$(PDF+\alpS)\unit{pb} assuming $\mtop = 172.5\GeV$.

{\tolerance=4000 The SUSY signal events are normalized to cross sections calculated at approximate NNLO+NNLL accuracy~\cite{Beenakker:1996ch,Beenakker:2016lwe,Kulesza:2008jb,Kulesza:2009kq,Beenakker:2009ha,Beenakker:2011fu,Borschensky:2014cia,Beenakker:2011sf,Beenakker:2010nq,Beenakker:2016gmf} obtained from the simplified model for the direct pair production of top squarks.
The cross sections of the \ttDM model are calculated at LO using \MADGRAPH~v2.4.2.\par}

\section{Event reconstruction}

In this section, the event reconstruction common to all the analyses presented in this paper is described.

An event may contain multiple primary vertices, corresponding to multiple $\Pp\Pp$ collisions occurring in the same bunch crossing. The candidate vertex with the largest value of summed physics-object $\pt^2$ is taken to be the primary $\Pp\Pp$ interaction vertex. The physics objects for this determination are the jets, clustered using the jet finding algorithm~\cite{Cacciari:2008gp,Cacciari:2011ma} using tracks assigned to candidate vertices as inputs, and the associated missing transverse momentum, taken as the negative vector sum of the transverse momentum of those jets. 

The particle-flow algorithm~\cite{bib:PF} aims to reconstruct and identify each individual particle in an event, with an optimized combination of information from the various elements of the CMS detector. The energy of photons is obtained from the ECAL measurement. The energy of electrons is determined from a combination of the electron momentum at the primary interaction vertex as determined by the tracker, the energy of the corresponding ECAL cluster, and the energy sum of all bremsstrahlung photons spatially compatible with originating from the electron track. The energy of muons is obtained from the curvature of the corresponding track. The energy of charged hadrons is determined from a combination of their momentum measured in the tracker and the matching ECAL and HCAL energy deposits, corrected for the response function of the calorimeters to hadronic showers. Finally, the energy of neutral hadrons is obtained from the corresponding corrected ECAL and HCAL energies.

For each event, hadronic jets are clustered from these reconstructed particles using the infrared and collinear safe anti-\kt algorithm~\cite{Cacciari:2008gp, Cacciari:2011ma} with a distance parameter of 0.4. The jet momentum is determined as the vectorial sum of all particle momenta in the jet, and is found from simulation to be, on average, within 5 to 10\% of the generated momentum over the whole \pt spectrum and detector acceptance. 

Additional $\Pp\Pp$ interactions within the same or nearby bunch crossings can contribute with additional tracks and calorimetric energy depositions to the jet momentum. To mitigate this effect, charged particles identified as originating from pileup vertices are discarded, and an offset correction is applied to correct for the contribution from neutral particles~\cite{Cacciari:2008pup}. Jet energy corrections are derived from simulation to bring the energy of a jet measured from the detector response to that of a particle-level jet on average. In situ measurements of the momentum balance in dijet, photon+jets, \zjets, and multijet events are used to account for any residual differences in jet energy scale between data and simulation~\cite{Khachatryan:2016kdb}. The jet energy resolution amounts typically to 15\% at 10\GeV, 8\% at 100\GeV, and 4\% at 1\TeV~\cite{Khachatryan:2016kdb}. Additional selection criteria are applied to each jet to remove jets potentially dominated by anomalous contributions from various subdetector components or reconstruction failures~\cite{CMS:2017wyc}. 
Jets produced by the hadronization of \PQb quarks are identified using \PQb tagging multivariate algorithms: DeepCSV~\cite{Sirunyan:2017ezt} for the inclusive searches and DeepJet~\cite{CMS-DP-2018-058, Bols:2020bkb} for the corridor search. The more recently developed DeepJet algorithm has slightly better performance for some parts of the phase space than the DeepCSV algorithm. All analyses use a medium working point for the tagger, corresponding to a a misidentification probability for jets originating from gluons or up, down, and strange quarks of 1\%, and a \PQb tagging efficiency of about 70\%. A tight working point, corresponding to a misidentification rate of 0.1\%, is also used in the analysis of Section~\ref{sec:s1l}.

The missing transverse momentum vector is defined as the negative vector \pt sum of all particle-flow candidates reconstructed in an event with jet energy corrections applied. Events with serious \ptmiss reconstruction failures are rejected using dedicated filters~\cite{Sirunyan:2019kia}.

The requirements imposed to select reconstructed particle objects specific to the separate search strategies incorporated into the present combination are given in the following sections. In Section~\ref{sec:bulk} we give brief summaries of the previously published searches, and in Section~\ref{sec:corridor} the detailed presentation of the new top quark corridor search.

\section{Inclusive top squark searches} 
\label{sec:bulk}

Three analyses targeting final states without leptons~\cite{Sirunyan:2021mrs}, with one~\cite{Sirunyan:2019glc}, or with two charged leptons~\cite{Sirunyan:2020tyy} have been previously published.
The main features are briefly discussed in this section.

\subsection{Fully hadronic analysis}

The search in the fully hadronic final state~\cite{Sirunyan:2021mrs} targets events with hadronic jets and large reconstructed \ptmiss.
The SM backgrounds with intrinsic \ptmiss generated through the leptonic decay of a \PW{} boson, where the neutrino escapes detection, are significantly suppressed by rejecting events containing isolated electrons or muons.
The contribution from events in which a \PW\ boson decays to a $\tau$ lepton is suppressed by rejecting events containing isolated hadronically decaying $\tau$ candidates~\cite{Khachatryan:2015dfa,CMS:2016gvn}.
A central feature of this analysis is the deployment of advanced jet tagging algorithms to identify hadronically decaying top quarks and \PW{} bosons, with different algorithms covering both the highly Lorentz-boosted regime and the resolved regime.
For the highly Lorentz-boosted regime, where the decay products of the particle in quest are expected to merge into a single large-$R$ jet with a distance parameter of $R = 0.8$, the \textsc{DeepAK8} algorithm~\cite{Sirunyan:2020lcu} is used to identify these large-$R$ jets originating from top quarks or \PW{} bosons.
In the resolved regime, where the decay products of the top quark are separately reconstructed using jets with $R = 0.4$, the \textsc{DeepResolved} algorithm~\cite{Sirunyan:2019glc} is used to tag these top quarks with intermediate \pt, ranging from 150 to 450\GeV.

To enhance sensitivity to signal models with a compressed mass spectrum where the mass of the top squark is close to the sum of the masses of the LSP and the \PW boson, a dedicated ``soft \PQb tag'' algorithm developed to identify very low \pt \cPqb hadrons is also used for the event categorization~\cite{CMS:2017kil}.
The analysis includes a total of 183 nonoverlapping signal regions, defined in Ref.~\cite{Sirunyan:2021mrs} and optimized for different SUSY models and ranges of mass splittings between SUSY particles.
A large \ptmiss, due to the presence of a pair of neutralinos in the signal model, is required.

The dominant sources of SM background with intrinsic \ptmiss are the inclusive production of top quark pairs, \PW and \PZ bosons, single top quark production, and the \ttbar{}\PZ process.
The contribution from \ttbar, \wjets, \ttbar{}\PW, and single top quark processes arises from events in which a \PW{} boson decays leptonically to produce \ptmiss associated with an energetic neutrino, but the charged lepton either falls outside of the kinematic acceptance or fails the lepton identification criteria.
This background is collectively referred to as ``lost-lepton'' background.
The contributions from \zjets and \ttbar{}\PZ events arise when the \PZ boson decays to neutrinos, resulting in large genuine \ptmiss.
Contributions from the QCD multijet process enter the search sample in cases where severe mismeasurements of jet momenta (\ie, jets passing through poorly instrumented regions of the detector) produce significant artificial \ptmiss, or when neutrinos arise from leptonic decays of heavy-flavor hadrons produced during the jet fragmentation.
The contribution of each SM background process to the search sample is estimated through measurements of event rates in dedicated background control samples that are translated to predicted event counts in the corresponding search sample with the aid of simulation.
The data are found to be in good agreement with the predicted backgrounds.

\subsection{Single-lepton analysis}\label{sec:s1l}

The search for top squark pair production in the single-lepton final state~\cite{Sirunyan:2019glc} focuses on final states with exactly 1 lepton, coming from the decay of a \PW{} boson from the decay chain $\PSQtDo\to\PQt\PSGczDo\to\PQb\PW\PSGczDo$ or $\PSQtDo\to\PQb\PSGcpmDo \to\PQb\PW\PSGczDo$.
Since the \PSGczDo in the final state of the signal gives rise to substantial \ptmiss compared with SM processes, $\ptmiss>250\GeV$ is required.
The transverse mass computed from the lepton \ptvec and \ptvecmiss is required to be larger than $150\GeV$ to reduce the lepton+jets background from $\ttbar$ and \wjets processes, for which \mT has a natural cutoff at the \PW boson mass (\mW).

The dileptonic \ttbar process, where one of the leptons is lost, is the largest remaining SM background.
In these lost-lepton events \mT is not bound by \mW because of the additional \ptmiss arising from the presence of a second neutrino.
The modified topness ($\tmod$) variable, introduced in Ref.~\cite{Sirunyan:2019glc}, is a measure for the likelihood of a single lepton event to originate from dileptonic \ttbar and is used to introduce better discrimination against this background.

The dileptonic \ttbar background is estimated through a set of dedicated control regions that require two isolated leptons.
The second lepton is added to \ptmiss in the calculation of variables that depend on \ptmiss, \eg \mT and \tmod, to mimic the lost-lepton scenario.

The subleading SM background comes from the process of \wjets production, where the \PW{} boson decays leptonically.
While the single-lepton events from the \PW boson are largely suppressed by the \mT requirement, events where the \PW{} boson is produced off-shell can still enter the signal regions.
The requirement of at least one \PQb-tagged jet significantly reduces this type of background.
Events are further categorized in terms of the invariant mass of the lepton and the \PQb-tagged jet, which helps to further reduce the \wjets background.
The \wjets background is estimated using control regions with an inverted \PQb-tagged jet requirement which yields a high-purity sample of \wjets events.

Irreducible SM backgrounds arise from the \ttZ and \PW{}\PZ{} processes when the \PZ{} boson decays into a pair of neutrinos.
These rare backgrounds and the remaining events from the single lepton \ttbar process are sub-dominant contributions in most search regions and are estimated using simulated samples.

This analysis also makes use of the same jet tagging algorithms, described above in the fully hadronic channel, to identify hadronic top quark decays in the final state.
This is motivated by the fact that none of the leading SM backgrounds, except \ttZ, has a hadronically decaying top quark in the final state, while in some signal scenarios one hadronically decaying top quark is expected.
Events in the lower \ptmiss search regions are categorized into different regions according to the presence of at least one merged or resolved top quark candidate.

Finally, a dedicated search strategy is used for signal scenarios with small mass splitting between the top squark and the LSP to optimize sensitivity.
In these compressed scenarios with $\Delta m\left(\PSQtDo,\PSGczDo\right)$ close to \mW or $m_{\PQt}$, \ptmiss can be small when neutralinos are back-to-back, and therefore \tmod and the merged and resolved top quark tags are not used.
Instead, one non-\PQb-tagged jet, which could arise from ISR for a signal event, is required and a requirement on the proximity of the lepton to the \ptmiss is introduced.
In the case of $\Delta m\left(\PSQtDo,\PSGczDo\right)\sim m_{\PW}$ at least one ``soft \PQb tag'', such as a secondary vertex, is required instead of the standard \PQb-tagged jets, to improve the acceptance for \PQb quarks that do not carry sufficient momentum to be reconstructed as a jet. 
In order to enhance the sensitivity to different signal scenarios events are categorized into 39 non-overlapping signal regions based on the values of \ptmiss and several of the variables introduced above.

\subsection{Dilepton analysis}\label{sec:dileptonInclusive}

The search in the dilepton final state~\cite{Sirunyan:2020tyy} is carried out using events containing a pair of leptons (electron or muons) with opposite charges.
The invariant mass of the lepton pair (\mll) is required to be greater than 20\GeV to suppress backgrounds with misidentified or nonprompt leptons from the hadronization of heavy-flavor jets in multijet events.
Events with additional leptons, including candidates with looser requirements on transverse momentum, and isolation are rejected.
Events with a same-flavor lepton pair that is consistent with the SM DY production are removed by requiring $\abs{m_{\PZ} - \mll} > 15\GeV$, where $m_{\PZ}$ is the mass of the \PZ boson.
To further suppress DY and other vector boson backgrounds, the number of jets is required to be at least two and, among them, the number of \PQb-tagged jets to be at least one.

The \ptmiss significance, denoted as \ptmissSig, is used to suppress events where detector effects and misreconstruction of particles from pileup interactions are the main source of reconstructed \ptmiss.
The algorithm used to obtain \ptmissSig is described in Ref.~\cite{Sirunyan:2019kia}.
A requirement of $\ptmissSig > 12$ is used in order to suppress the otherwise overwhelming DY background in the same-flavor channel.
This requirement exploits the stability of \ptmissSig with respect to the pileup rate for events with no genuine \ptmiss.
The DY background is further reduced through a requirement on the azimuthal angular separation between \ptvecmiss and the momentum of the leading~(subleading) jet of $\cos\Delta\phi(\ptmiss, \mathrm{j}) < 0.80\,(0.96)$.
These criteria reject a small background of DY events with significantly mismeasured jets.

The main variable in this analysis is \mtll, which is defined in equation \eqref{eq:mtll}, and extensively discussed in Ref.~\cite{Sirunyan:2017leh}.
The key feature of the \mtll observable is that it retains a kinematic endpoint at the \PW boson mass for background events from the leptonic decays of two \PW bosons, produced directly or through a top quark decay.
Similarly, the \mtlblb observable, defined with equation~\eqref{eq:mtll}, but using the vector sum of the leptons and the \cPqb-jets instead of leptons alone~\cite{Sirunyan:2020tyy}, is bounded by the top quark mass if the leptons, neutrinos and \PQb-tagged jets originate from the decay of top quarks.
By contrast, signal events do not have the respective endpoints and are expected to populate the tails of these distributions.

Signal regions based on \mtll, \mtlblb, and \ptmissSig are defined to enhance the sensitivity to different signal scenarios.
The regions are further divided into different categories separately for events with a same-flavor and a different-flavor lepton pair, to account for the different SM background composition.
The signal regions are defined such that there is no overlap between them, nor with the background-enriched control regions.

Events with an opposite-charge lepton pair are abundantly produced by the DY and \ttbar processes.
The event selection rejects the vast majority of DY events.
Therefore, the major backgrounds from SM processes in the search regions are top quark pair and single top events that pass the \mtll threshold because of severely mismeasured \ptmiss or a misidentified lepton.
In high \mtll and \ptmissSig signal regions, \ttZ events with $\PZ\to\cPgn\cPagn$ are the main SM background.
Remaining DY events with large \ptmiss from mismeasurement, multiboson production and other \ttbar/single~\PQt processes in association with a \PW, a \PZ or a Higgs boson (\ttW, \tqZ, or \ttH) are sources of smaller contributions.
A detailed description of the background estimation method is given in Ref.~\cite{Sirunyan:2020tyy}.

\section{Top quark corridor analysis}\label{sec:corridor}

The top quark corridor analysis is discussed in this section in more detail, as it is presented for the first time in this paper. In this search, events containing a dilepton pair with opposite charge and \ptmiss are selected, and a DNN algorithm is used to increase the sensitivity to the signal. The full DNN score distribution for events in the signal region is used to test the presence of the signal.

\subsection{Object and event selection}
The object selection and baseline requirements of the event selection are the same as those for the dilepton analysis summarized in the first paragraph of Section~\ref{sec:dileptonInclusive}, and are detailed in this section. Electron and muon candidates are required to have $\pt > 20\GeV$ and $\abs{\eta} < 2.4$. In addition, the \pt of the leading lepton must be at least 25\GeV. The leptons are required to be isolated by measuring their relative isolation as the scalar \pt sum, normalized to the lepton \pt, of the photons and of the neutral and charged hadrons within a cone of radius $\Delta R=\sqrt{\smash[b]{(\Delta\eta)^2+(\Delta\phi)^2}}=0.3$ (0.4) around the candidate electron (muon). In order to reduce the dependence on the number of pileup interactions, charged hadron candidates are included in the sum only if they are consistent with originating from the selected primary vertex in the event. The expected contribution from neutral hadrons due to pileup is estimated following the method described in Ref.~\cite{Khachatryan:2015hwa}. For an electron candidate the relative isolation requirement depends on $\eta$ (values close to 0.04) and for a muon it is required to be smaller than 0.15.

Selected jets are required to have $\pt > 30\GeV$ and $\abs{\eta} < 2.4$. Additionally, jets that are found within a cone of $\Delta R=0.4$  around the selected leptons are rejected. Jets originating from the hadronization of bottom quarks are identified as \cPqb-tagged jets by using the medium working point of the DeepJet algorithm~\cite{CMS-DP-2018-058, Bols:2020bkb}.

Simulated events are corrected to account for differences with respect to data in the lepton reconstruction, identification, and isolation efficiencies, as well as efficiencies in the performance of \cPqb tagging. The values of the data-to-simulation correction factors are parameterized as functions of the \pt and $\eta$ of the object and deviate from unity by less than 1\% for leptons and less than 10\% for \cPqb-tagged jets.

Selected events are classified in categories according to the flavor of the two leading leptons (\ee, \emu, \mumu) and the data-taking period (2016, 2017, 2018). Moreover, events are required to contain at least two jets, one of which must be \cPqb tagged. This set of requirements is referred to as the baseline selection.

After the baseline selection, most of the background events (about 98\%) are expected to come from \ttbar, \tW, and DY processes. To suppress these backgrounds, the signal region is defined with the requirements $\ptmiss>50\GeV$ and $\mtll > 80\GeV$.  As described in Sec.~\ref{sec:dileptonInclusive}, $\mtll$ serves to account for the multiple sources of \ptmiss in the signal process and to exploit the differences with respect to the background processes. For \ttbar, \tW or \wjets events this variable's distribution has a kinematic endpoint at the \PW boson mass, because the transverse mass of each lepton-neutrino pair corresponds to the transverse mass of the \PW boson, whereas signal events have neutralinos contributing to the total \ptmiss, so they populate larger $\mtll$.

\subsection{Background estimation}\label{sec:corridorbkg}

Although most of the \ttbar events are rejected by requiring $\mtll > 80\GeV$, it is still the dominant background contribution in the signal region, where most of the events have a large \mtll value because of resolution effects when computing \ptvecmiss. In this region, some of the \ttbar events contain jets with a mismeasured energy and, in a smaller proportion, there are events where one of the leptons is missed and a lepton that is not from a \PW boson decay (nonprompt lepton) is taken as the second lepton in the event. The effect of the jet mismeasurements is checked in MC and an uncertainty is assigned. Events containing nonprompt leptons are considered in a different background category. 

The second-largest contribution is \tW~background, which is approximately 4\% of the total background, and is also estimated from MC simulation. The DY events give the third-largest background contribution in the same-flavor channel, while  their contribution is negligible in the \emu channel. 

Background with nonprompt leptons is estimated from MC simulation and validated in a control region with the same selection as the signal region, but requiring two same-charge leptons. These events include the contribution from jets misidentified as leptons or with leptons coming from the decay of a bottom quark mistakenly identified as coming from the hard process. In the same-charge region, most of the events come from \ttbar with nonprompt leptons, with a smaller contribution of events with prompt leptons from \ttW and \ttZ~production, and dileptonic \ttbar with prompt leptons and a mismeasurement of the electron charge. A reasonable agreement with same-charge data, within 10--15\%, is observed in this validation region.
Minor background contributions are also estimated from MC simulation and come from diboson ($\PW\PW$, $\PW\PZ$, and $\PZ\PZ$), \ttW, and \ttZ~events, with a total contribution of about 1\%.

\begin{figure*}[htb!]
\centering
\includegraphics[width=1\cmsFigWidth]{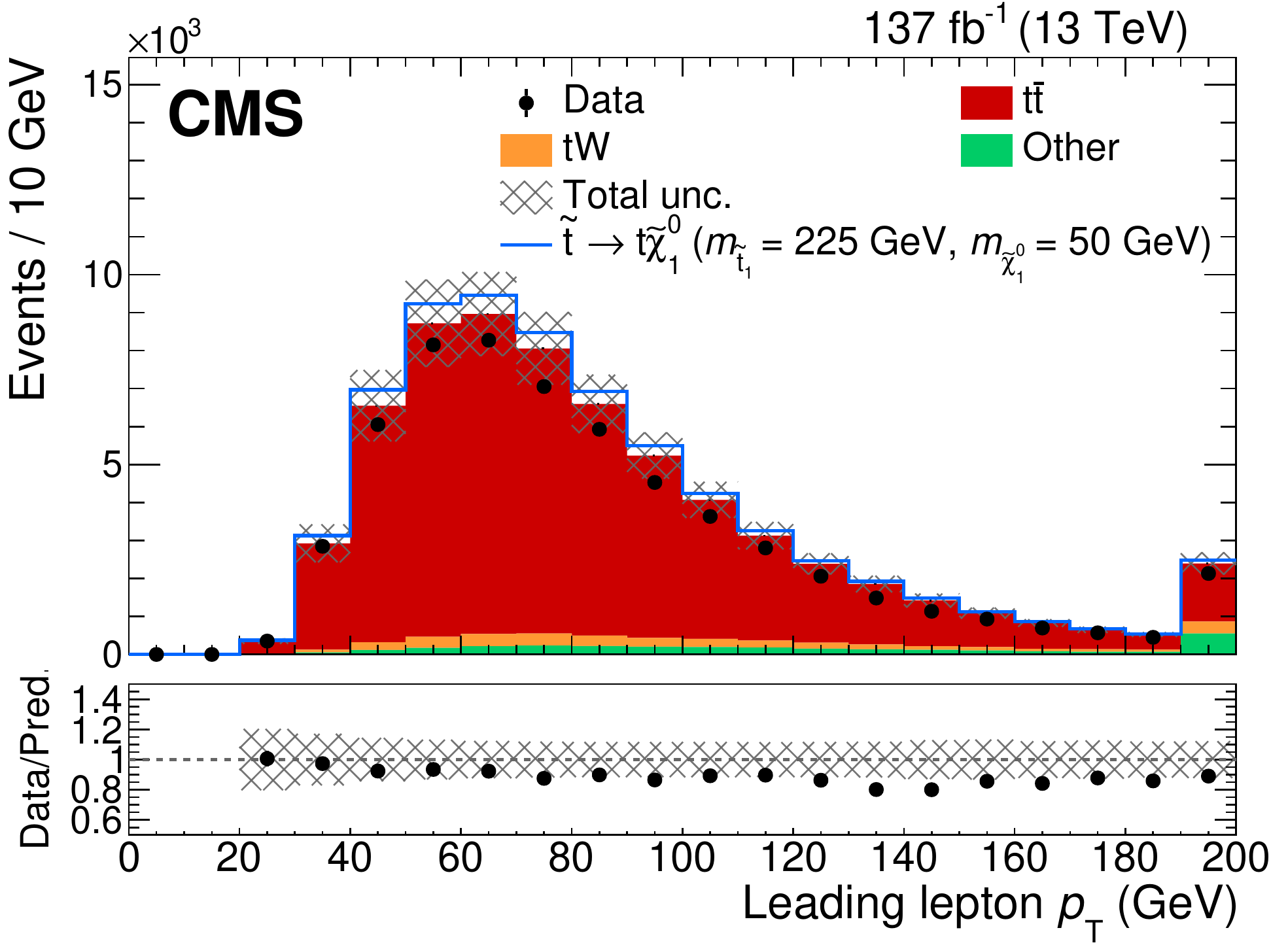}
\includegraphics[width=1\cmsFigWidth]{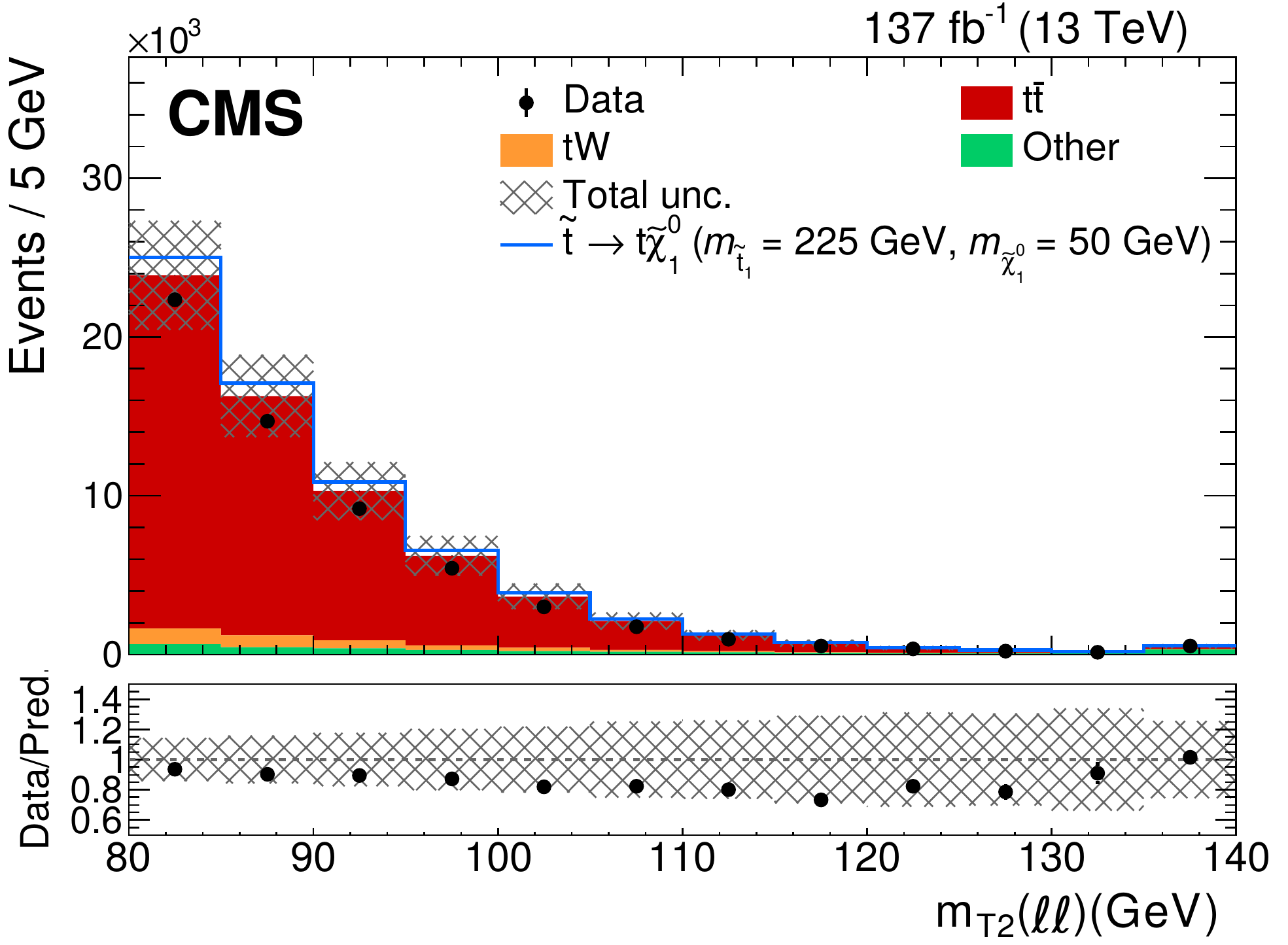}
\includegraphics[width=1\cmsFigWidth]{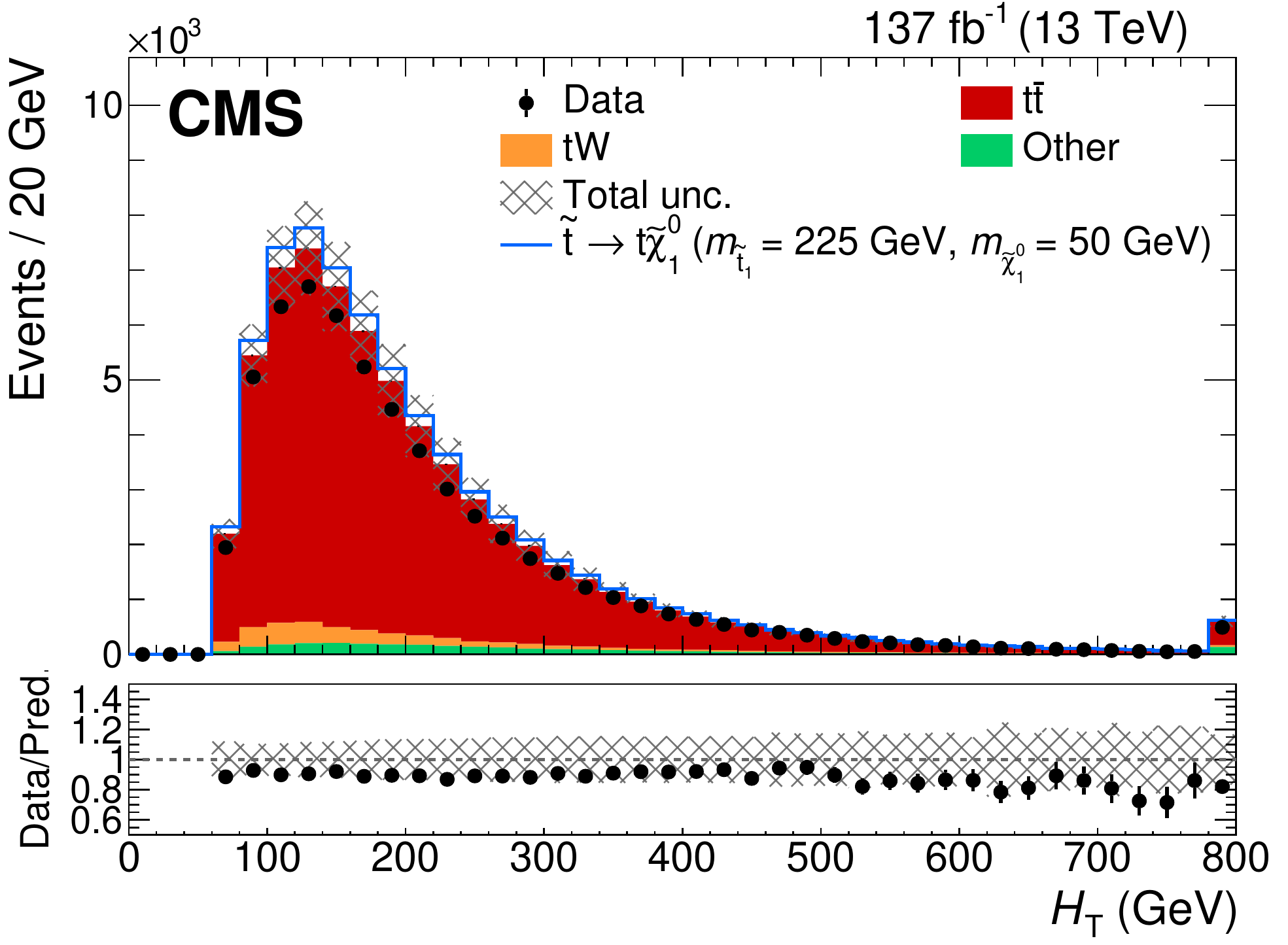}
\includegraphics[width=1\cmsFigWidth]{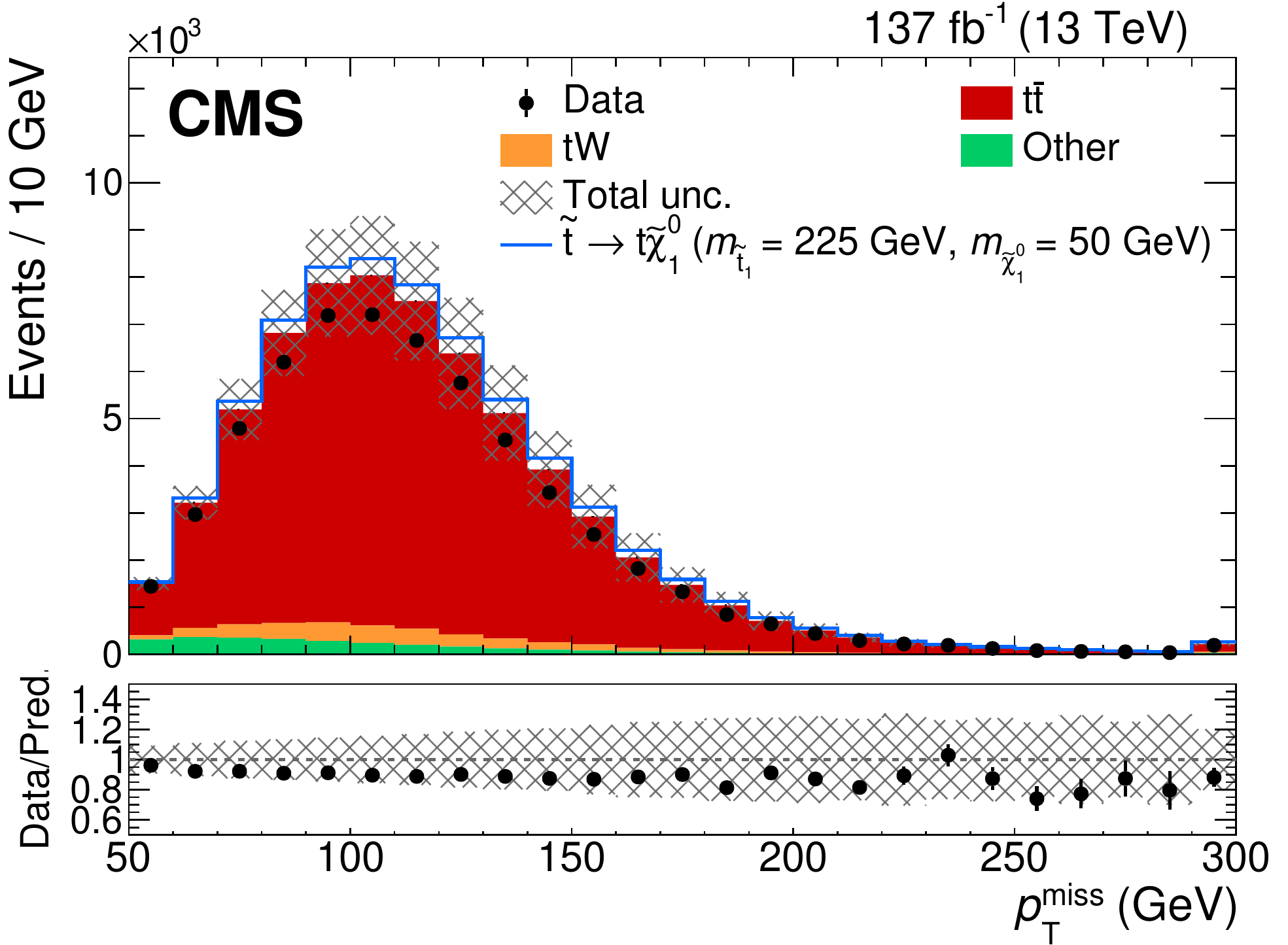}
\caption{Pre-fit distributions of data and MC events in the signal region with the signal stacked on above the background prediction for a mass hypothesis of $\mstop = 225\GeV$ and $\mlsp = 50\GeV$. Events from $\ttbar\PW$, $\ttbar\PZ$, DY, nonprompt leptons, and diboson processes are grouped into the 'Other' category. The lower panel contains the data-to-SM prediction ratio. The uncertainty band includes statistical, background normalization and all systematic uncertainties described in Section~\ref{sec:unc}. From upper left to lower right: leading lepton $\pt$, $\mtll$, \HT, and $\ptmiss$.}
\label{fig:dataMCplotsSR}
\end{figure*}

The distributions of the main observables in data, the leading lepton \pt, \mtll, the scalar sum of the \pt of all the selected jets (\HT) and \ptmiss in the signal region, are shown in Fig.~\ref{fig:dataMCplotsSR}. The simulation and data are generally in agreement within the uncertainties. The uncertainties are described in Section~\ref{sec:unc}.

\subsection{Search strategy}\label{strategy}

In order to maximize the sensitivity and to exploit all the differences between the signal and \ttbar background, a multivariate analysis is implemented using a DNN, trained with events passing the baseline selection. The DNN takes into account all the shape differences between signal and background distributions for the training variables, including correlations, in turn achieving a strong final discriminator. The signal model used was the direct pair production of top squarks, for a sequence of \mstop mass values in the range 145--295\GeV and \deltamcor ranging from 0 to 30\GeV. The background input to the training was simulated \ttbar with \emu decays. To avoid overfitting, 40\% of the total \ttbar and signal events are used for the training and the rest for the signal extraction.

The training was done using events passing the baseline selection in order to use the separation power of different observables over a large range. A total of 13 variables are selected for the training: top squark and neutralino masses (\mstop, \mlsp), the transverse momentum of the electron-muon pair ($\pt^{\Pe\PGm}$), the angle between the momentum of the leptons in the transverse plane ($\Delta\phi(\Pe\PGm)$), the pseudorapidity difference between the leptons ($\Delta\eta(\Pe\PGm)$), the momenta and pseudorapidities of the individual leptons, \mll, \mtll, \ptmiss, and \HT.

\begin{figure*}[htb!]
\centering
\includegraphics[width=1\cmsFigWidth]{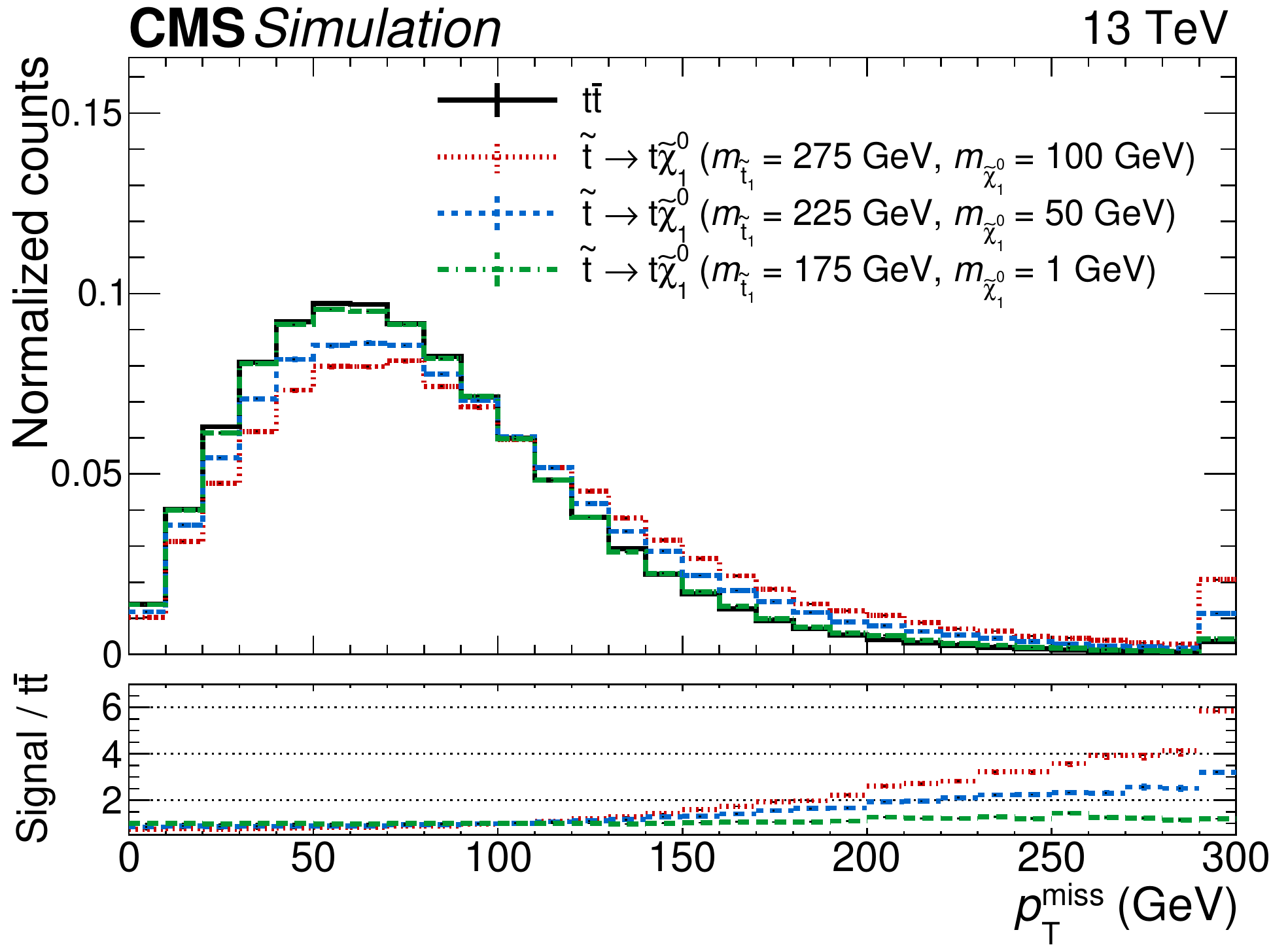}
\includegraphics[width=1\cmsFigWidth]{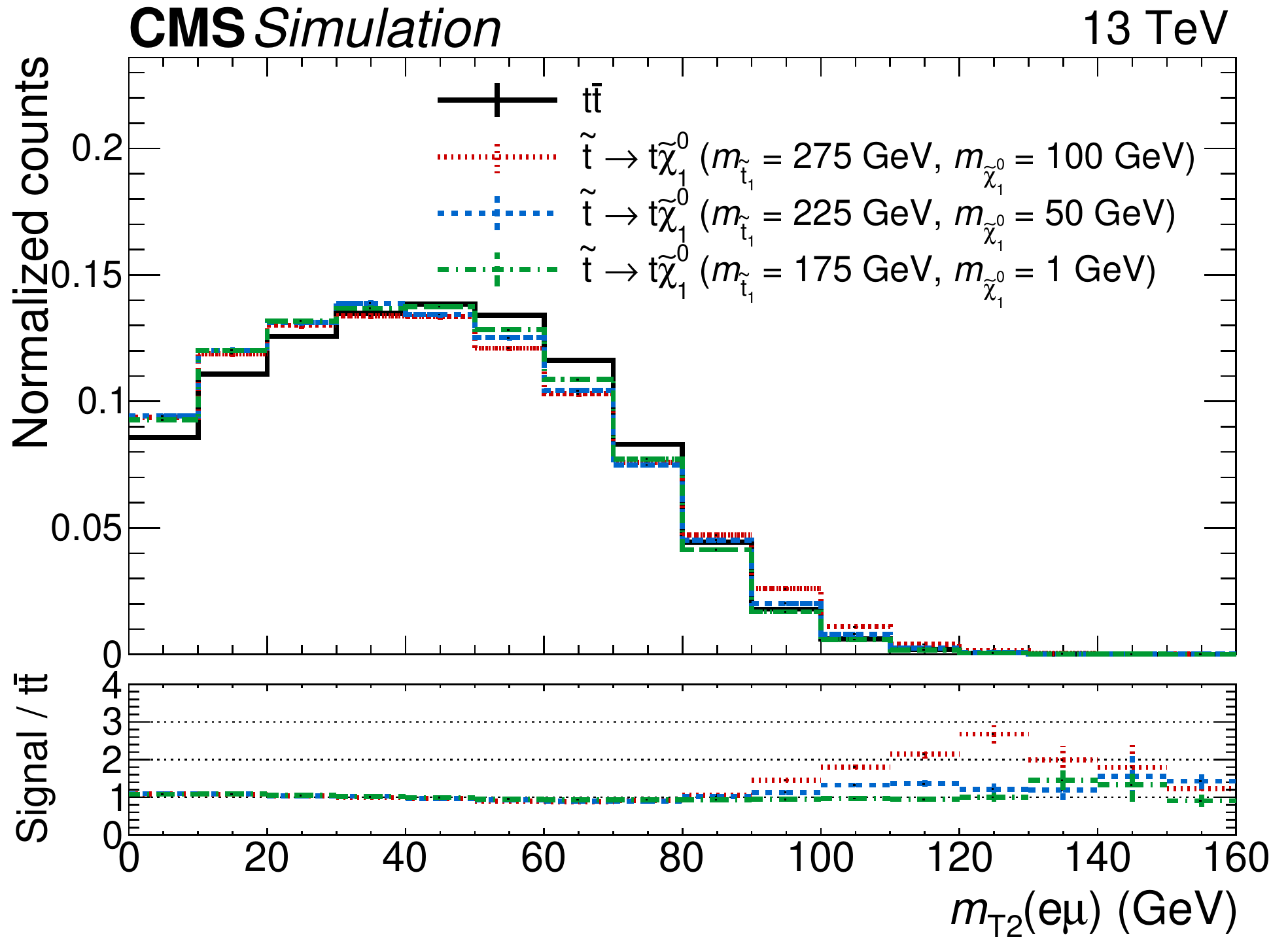}
\includegraphics[width=1\cmsFigWidth]{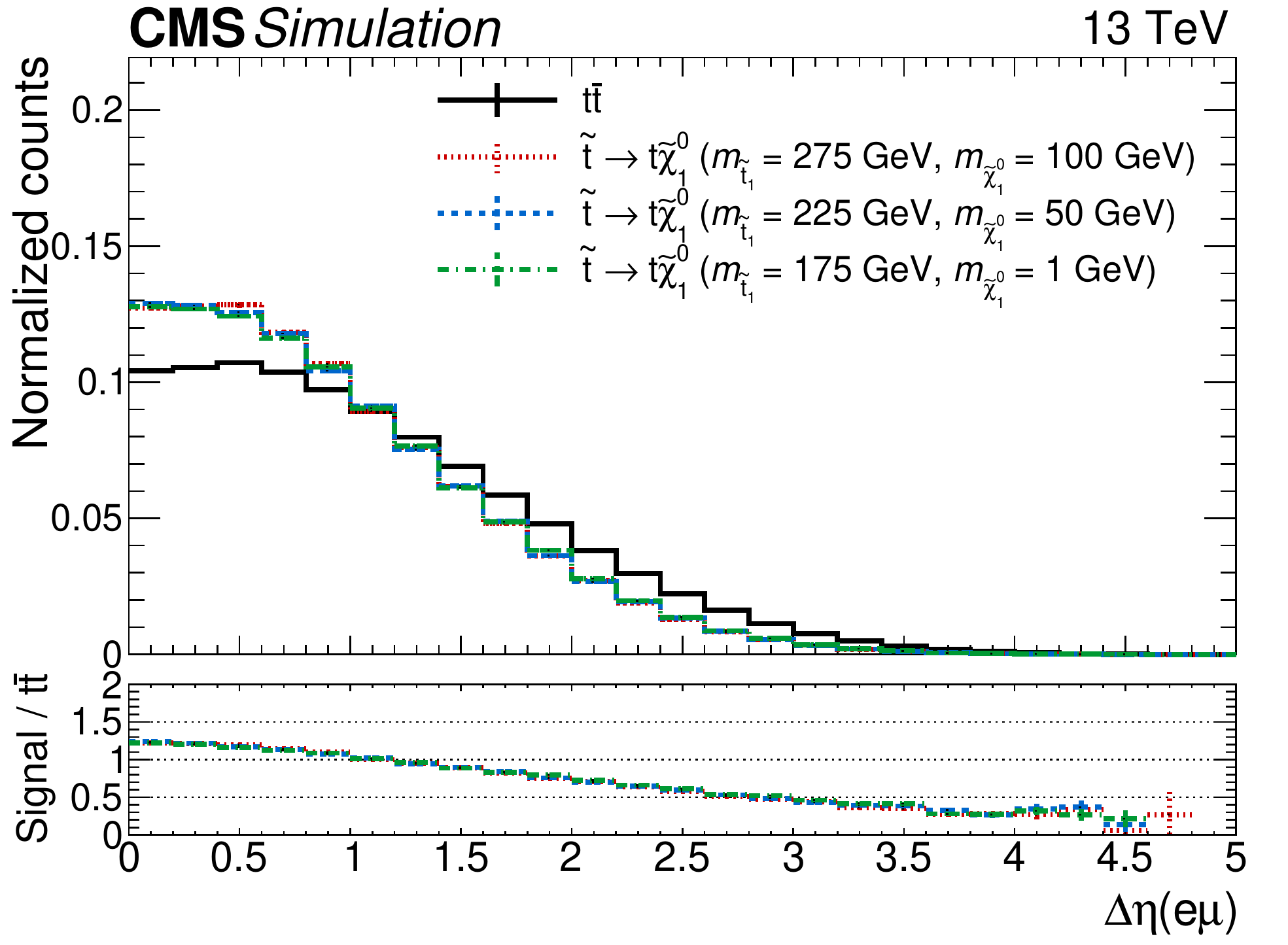}
\includegraphics[width=1\cmsFigWidth]{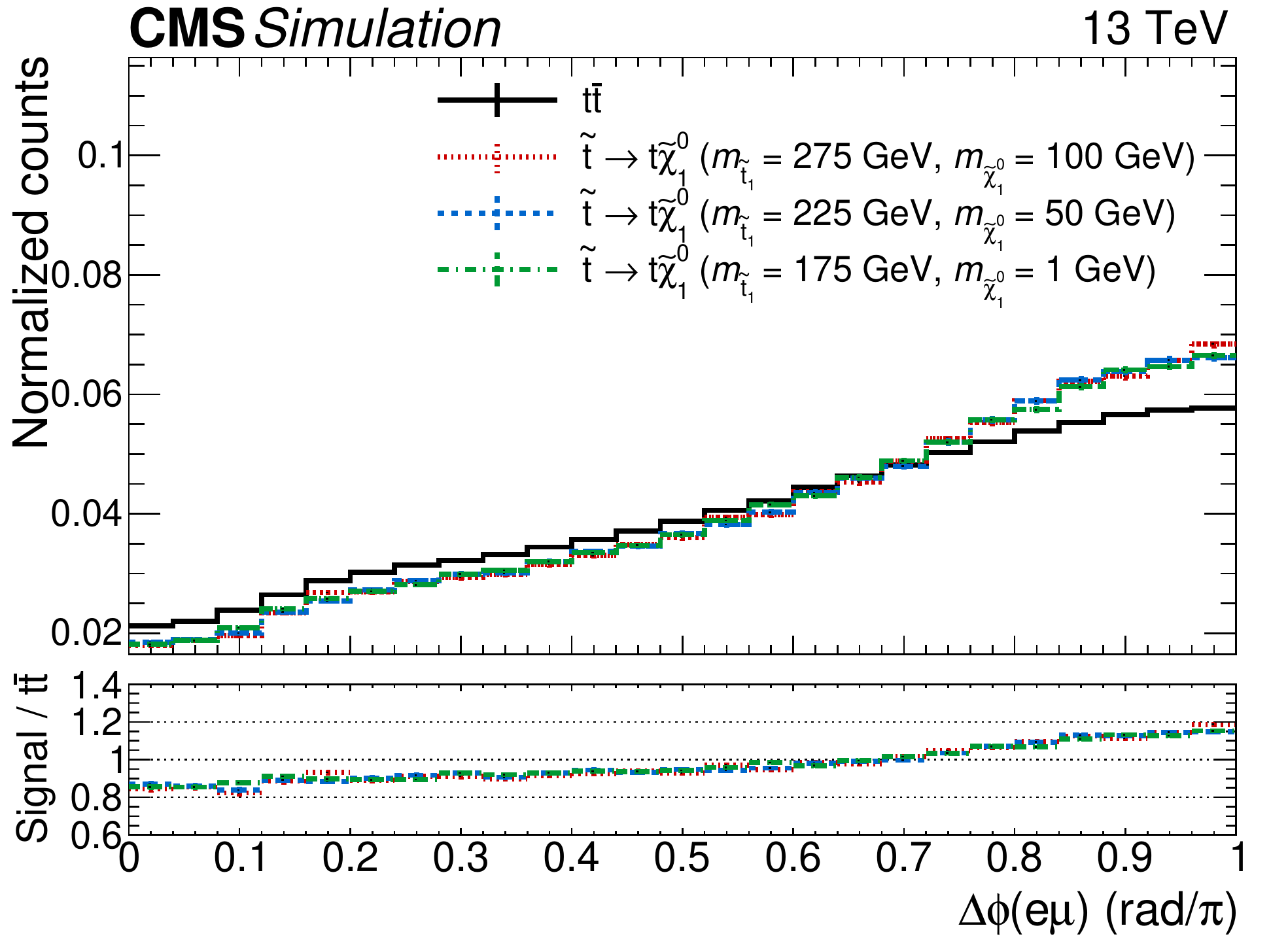}
\caption{Normalized distributions for some of the training variables in the baseline selection. Distributions for signal points with different top squark and neutralino masses and SM \ttbar events are compared. From upper left to lower right: \ptmiss, $\mTii(\emu)$, $\Delta\eta(\emu)$, and $\Delta\phi(\emu)$.}
\label{fig:BScomp}
\end{figure*}

Figure \ref{fig:BScomp} shows the normalized distributions of the most discriminating variables for \ttbar and signal samples for different values of \mstop and \mchi~, after the baseline selection. This figure also shows that, in some variables, the shape of the distributions does not have the same behavior for all the signal points. The differences in \ptmiss and $\mtll$ between signal and background are larger for signal points with large \mchi. To exploit these differences and improve the sensitivity, a parametric DNN~\cite{Baldi:2016fzo} is used, in which the top squark and neutralino masses are introduced in the training. In this way, a specific model for each signal point training a single DNN is achieved. For background events, \mstop and \mchi~are randomly taken, to avoid introducing correlations, using a probability distribution that matches the values of \mstop~and \mchi~in the signal sample.

The training was performed with  \textsc{TensorFlow}~\cite{tensorflow2015-whitepaper} using the \textsc{Keras} interface~\cite{chollet2015keras}. All the hyperparameters are optimized with the aim of avoiding overfitting and achieving the highest possible accuracy on the classification. The final DNN structure is sequential: 7 hidden layers with a ReLU activation function~\cite{chollet2015keras} ($300, 200, 100, 100, 100, 100, 10$ neurons). The output consists of two neurons with a softmax normalization function~\cite{chollet2015keras}, which allows one to interpret the output numbers as probabilities. The optimizer that is selected corresponds to Adam~\cite{kingma2014adam} with a learning rate of 0.0001. Out of the 40\% of events used for the DNN implementation, 60\% are used for training, 15\% for validation, and the rest to check that the DNN works properly and there is no overfitting.

\begin{figure*}[h!]
\centering
\includegraphics[width=0.6\textwidth]{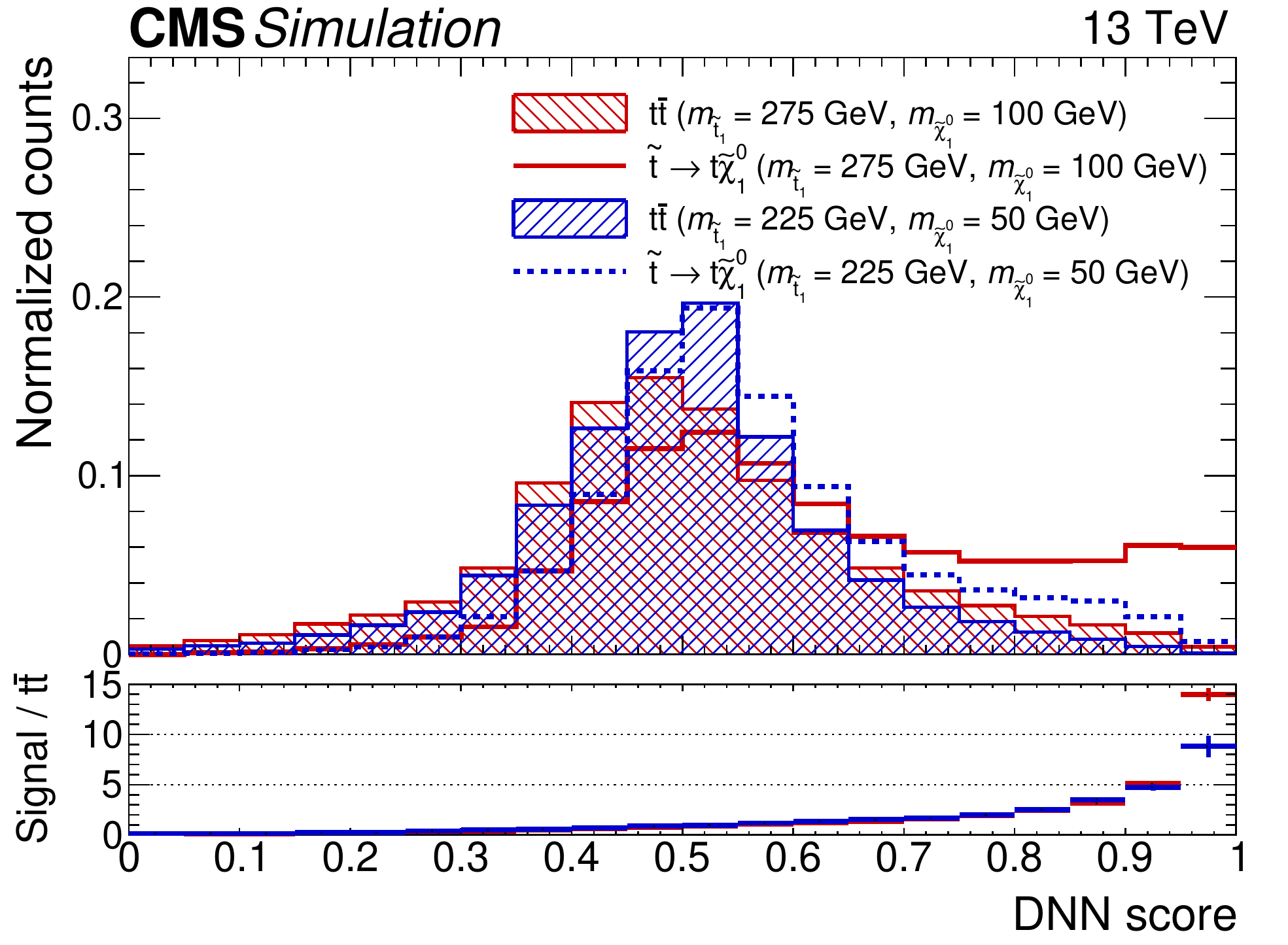}
\caption{\label{fig:dnnoutput}Normalized DNN score distribution comparing the signal and the \ttbar~background in the signal region for two mass hypotheses: $\mlsp=$ 50 (100)\GeV and $\mstop=$ 225 (275)\GeV.}
\end{figure*}

Figure \ref{fig:dnnoutput} shows the DNN output  for two different mass parameters in the signal region for signal and \ttbar background. Since both masses are introduced in the training, the DNN score shape is different for both signal and background. This figure shows that the DNN score is a good discriminator between signal and background, especially at high values of the distribution.

The gain in sensitivity by using the DNN score instead of using only the \ptmiss distribution increases with increasing \deltamcor and with increasing $\mlsp$ for a fixed \deltamcor. For the fully degenerate case ($\mstop = 175\GeV$, $\mlsp = 1\GeV$) the kinematics of the SUSY process are very similar to the \ttbar background, so using the DNN does not help to improve the separation. The sensitivity to that point relies completely on the total measured cross section. For larger \mstop and \mlsp, even if $\deltamcor = 0$, the DNN starts to improve the sensitivity (as shown in Fig. \ref{fig:dnnoutput}). The score shape separation between signal and background also starts to increase for relatively low \mstop and $\mlsp$ if $\deltamcor > 0$.

The modeling of the DNN is checked in a validation region in which the signal region selection requirements are applied, except that $\ptmiss < 50\GeV$ and $\mtll < 80\GeV$ are required, and that only the \emu channel is used. This region is orthogonal to the signal region, and the signal contamination is expected to be small for the signal masses in which the sensitivity relies on the DNN discriminant. This region is highly dominated by \ttbar and \tW events and a good agreement with data is observed. 
Furthermore, the DY modeling of the DNN output distribution is also checked in a validation region where the invariant mass of the same-flavor lepton pairs is close to the mass of the $\PZ$ boson. The DY background is observed to be well modeled and populates preferentially low DNN score bins.

\subsection{Systematic uncertainties} \label{sec:unc}

Several sources of systematic uncertainty affect the background prediction and signal yields. Modeling of the trigger, efficiencies of the lepton reconstruction, identification and isolation, the jet energy scale and resolution, efficiencies of the \cPqb~tagging and mistag rate, and the pileup modeling have uncertainties that are considered in the estimate of both background and signal yields. All these sources are described in Section \ref{sec:experimentalsyst}. 

As the \ttbar background plays an essential role and needs to be accurately estimated, various modeling uncertainties are taken into account. These uncertainties consider variations of the main theoretical parameters used in the simulation and have been studied previously by the CMS Collaboration \cite{CMS-PAS-TOP-16-021,Sirunyan:2018ptc}. These uncertainties are explained in detail in Section \ref{sec:ttmodelling}.

Uncertainties in signal modeling are described in Section~\ref{sec:uncsignalmod}. Section~\ref{sec:otherunc} includes other sources of uncertainty as the background and signal normalization uncertainties.

\subsubsection{Experimental uncertainties}\label{sec:experimentalsyst}

The following experimental uncertainties are calculated for every background and signal estimate and are propagated to the final DNN output distribution in the signal region.

The uncertainties in the trigger, lepton identification, and isolation efficiencies used in simulation are estimated by varying data-to-simulation scale factors by their uncertainties, which are about 1.5\% for electron identification and isolation efficiencies, 1\% for muon identification and isolation efficiencies, and about 1.5\% for the trigger efficiency. The uncertainties in the muon momentum scales are taken into account by varying the momenta of the muons by their uncertainties, taken from the muon momentum scale corrections \cite{Sirunyan:2018fpa}. All these uncertainties are considered as correlated between years. 

For the \cPqb~tagging efficiency and mistag rate the uncertainties are determined by varying the scale factors for the \cPqb-tagged jets and mistagged light-flavor quark and gluon jets, according to their uncertainties, as measured in QCD multijet events~\cite{CMS-DP-2018-058, Bols:2020bkb,Sirunyan:2017ezt}. The uncertainties related to the jet energy scale and jet energy resolution are calculated by varying these quantities in bins of \pt and $\eta$, according to the uncertainties in the jet energy corrections, which amount to a few percent~\cite{Khachatryan:2016kdb}. The uncertainty in the effect of the jet mismeasurements, described in Section~\ref{sec:corridorbkg}, is added to the jet energy resolution uncertainties. This uncertainty is taken as partially correlated between years. 

The uncertainty in \ptmiss~from the contribution of unclustered energy is evaluated based on the momentum resolution of the different particle-flow candidates, according to their classification. Details on the procedure can be found in Refs.~\cite{bib:PF,CMS:EGM-14-001,TRK-11-001}. The uncertainty in the modeling of the contributions from pileup collisions is evaluated by varying the inelastic {$\Pp\Pp$}~cross section in the simulation by $\pm4.6$\%~\cite{Sirunyan:2018nqx}. These uncertainties are treated as correlated between data periods.

A summary of the experimental uncertainties in the \ttbar background and signal is shown in Table \ref{tab:expunc}. These uncertainties are also applied to the prediction of other minor backgrounds and have an effect in both the shape and the normalization.

\begin{table}[htb!]
\centering \topcaption{Summary of the contributions of the experimental uncertainties in the DNN score distribution for signal and the \ttbar background. The values represent the relative variation in the number of expected events across different signal models in the signal region.}
\begin{tabular}
{l c c c}
{Source} &  \multicolumn{3}{c}{Uncertainties (\%)} \\
	{} &  { \ttbar } & & signal \\
\hline
Electron efficiency      & & {1.5 } & \\
Muon efficiency          & &{0.5 } &\\
Trigger modeling         & &{1.2 } &\\
Muon energy scale        & &{1.4 } &\\
\cPqb~tagging efficiency & &{3.0 } &\\
Jet energy resolution    & 16.0 & &7.0\\
Jet energy scale         & 7.5  & &5.7\\
Unclustered energy       & 4.2 & &5.0 \\
Pileup modeling          & 3.2 & &1.5 \\
Size of the MC sample    & 3.0 & &25.0 \\
\end{tabular}
\label{tab:expunc}
\end{table}

\subsubsection{Modeling uncertainties in the  \texorpdfstring{\ttbar}{ttbar} background}\label{sec:ttmodelling}

Modeling uncertainties for the \ttbar background are calculated by varying different theoretical parameters in the MC generator within their uncertainties and propagating their effect to the final distributions.

The uncertainty in the modeling of the hard-interaction process is assessed in the \POWHEG sample varying up and down $\mu_\mathrm{F}$ and $\mu_\mathrm{R}$ by factors of 2 and 1/2 relative to their common nominal value of $\mu_\mathrm{F}^2 = \mu_\mathrm{R}^{2} = \mtop^2 + p^2_{\mathrm{T},\PQt}$. Here $p_{\mathrm{T},\PQt}$ denotes the \pt of the top quark in the \ttbar rest frame. The effect of this variation is propagated to the acceptance and efficiency, estimated using the \ttbar~\POWHEG sample. This uncertainty is correlated among the data-taking periods.

The uncertainty in the choice of the PDFs and in the value of $\alpS$ is determined by reweighting the sample of simulated \ttbar events according to the envelope of a PDF set of 100 NNPDF3.0 replicas~\cite{Ball:2014uwa} for 2016 and 32 PDF4LHC replicas~\cite{Butterworth:2015oua} for 2017 and 2018. 
The uncertainty in $\alpS$ is propagated to the acceptance by reweighting the simulated sample by sets of weights with two variations within the uncertainties of $\alpS$. Only the uncertainties for the 2017 and 2018 periods are taken to be correlated, while the 2016 period is kept uncorrelated, because the PDF set used is different.

The effect of the modeling uncertainties of the initial-state and final-state radiation is evaluated by varying the parton shower scales (running \alpS) by factors of 2 and 1/2~\cite{bib:powheg2}. In addition, the impact of the matrix element and parton shower matching, which is parameterized by the \POWHEG generator as $h_{\mathrm{damp}} = 1.58^{+0.66}_{-0.59}~\mtop$~\cite{CMS-PAS-TOP-16-021,Sirunyan:2019dfx}, is calculated by varying this parameter within the uncertainties. This uncertainty is calculated using dedicated \ttbar samples and is taken as correlated between the years.

To model the measured underlying event the parameters of \PYTHIA~are tuned~\cite{Skands:2014pea,Sirunyan:2019dfx}. An uncertainty is assigned by varying these parameters within their uncertainties using dedicated \ttbar samples. The uncertainty corresponding to the 2016 period is applied for the CUETP8M2T4 tune and is kept as uncorrelated to the uncertainty on the CP5 tune for 2017 and 2018, which is fully correlated for the two periods.

An uncertainty on the \pt of the top quark is also considered to account for the known mismodeling found in the \POWHEG MC sample~\cite{CMS-PAS-TOP-16-021}. A reweighting procedure exists to fix the mismodeling but, to avoid biasing the search, we use unweighted distributions and assign an uncertainty from the full difference to the weighted distributions.

For the top quark mass, 1\GeV is conservatively taken as the uncertainty, which corresponds to twice the uncertainty of the CMS measurement~\cite{Khachatryan:2015hba}, and is also propagated to the acceptance. The differences in the final yields for each bin of the DNN score distribution between the \ttbar background prediction with $\mtop = 172.5 \pm 1.0\GeV$ are taken as an uncertainty, accounting for the possible bias introduced in the choice of $\mtop = 172.5\GeV$ in the MC simulation. The uncertainty is assessed using dedicated \ttbar samples produced with a different \mtop.

The modeling uncertainties in the signal region yields for the \ttbar background are summarized in Table \ref{tab:ttbarunc}; they have an effect on the shape and also on the normalization.

\begin{table*}[htb!]
\centering \topcaption{Summary of the contribution of each modeling uncertainty source to the DNN score distribution for the \ttbar background.}
\begin{tabular}
{l c}
{Source} &  Average for \ttbar(\%) \\
\hline
PDFs and $\alpS$ (acceptance) & 1.0 \\
$\mu_\mathrm{F}$, $\mu_\mathrm{R}$ scales (acceptance) & 3.8 \\
Initial-state radiation& 0.6 \\
Final-state radiation & 3.4 \\
Top \pt & 1.3\\
Matrix element/parton shower matching & 2.0 \\
Underlying event & 1.5 \\
Top quark mass (acceptance)& 1.5 \\
\end{tabular}
\label{tab:ttbarunc}
\end{table*}

\subsubsection{Signal modeling}\label{sec:uncsignalmod}
The effect on the signal model of the ISR reweighting, described in Section~\ref{sec:MC}, is considered. Half of the deviation from unity is taken as the systematic uncertainty in these reweighting factors. This uncertainty is propagated to the final distribution and taken as a shape uncertainty.

The uncertainty in the modeling of the hard interaction in the simulated signal sample is calculated varying up and down $\mu_\mathrm{F}$ and $\mu_\mathrm{R}$ by factors of 2 and 1/2 relative to their nominal value. In addition, a 6.5\% uncertainty in the signal normalization is assigned, according to the uncertainties in the predicted cross section of signal models in the top squark mass range of the analysis~\cite{Borschensky:2014cia}.

\subsubsection{Other uncertainties} \label{sec:otherunc}
The uncertainty in the overall integrated luminosity for the combined sample, which affects the signal and background normalization, amounts to 1.6\%~\cite{CMS-LUM-17-003, CMS-PAS-LUMI-2017, CMS-PAS-LUMI-17-004, CMS-PAS-LUMI-18-002}. The total uncertainty is split in different sources, partially correlated across years.

A normalization uncertainty is applied to each background and signal estimate separately. The uncertainty in the \ttbar normalization is taken from the uncertainty in the NNLO+NNLL cross section, as quoted in Section \ref{sec:MC}, and additionally the top quark mass is varied by $\pm 1\GeV$, leading to a variation of the cross section of 6\%.

For DY, dibosons, $\ttbar\PW$, and $\ttbar\PZ$ processes a 30\% normalization uncertainty is assigned covering the uncertainty in the theoretical cross section and in the measurements. For the \tW process an uncertainty of 12\% is assigned. In the case of the nonprompt lepton background, a normalization uncertainty of 30\% is also applied, covering the largest deviations observed in the same-charge control region described in Section~\ref{sec:corridorbkg}.

Statistical uncertainties arise from the limited size of the MC samples. They are considered for each signal and background process, in each bin of the distributions. They are introduced through the Barlow--Beeston approach~\cite{statisticalUnc}.

All the systematic uncertainties described in Sections~\ref{sec:experimentalsyst} and \ref{sec:ttmodelling} are assigned to each DNN distribution bin individually, and treated as correlated among all the bins and all processes. The statistical uncertainties are treated as uncorrelated nuisance parameters in each bin of the DNN score distribution. All of the systematic uncertainties are treated as fully correlated among the different final states.

\section{Results}

\subsection{Corridor results}
\begin{figure*}[htb!]
\centering
\includegraphics[width=1\cmsFigWidth]{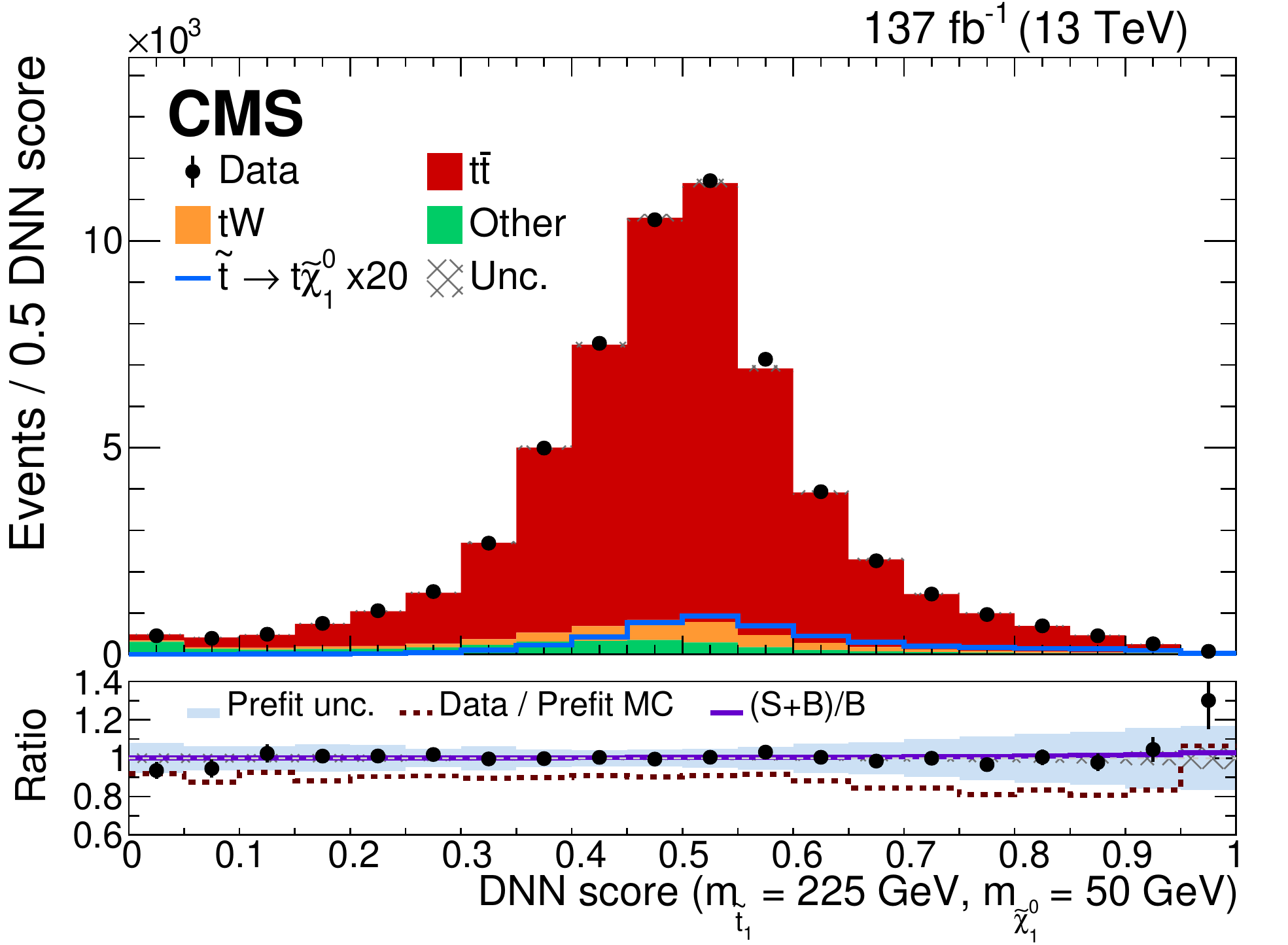}
\includegraphics[width=1\cmsFigWidth]{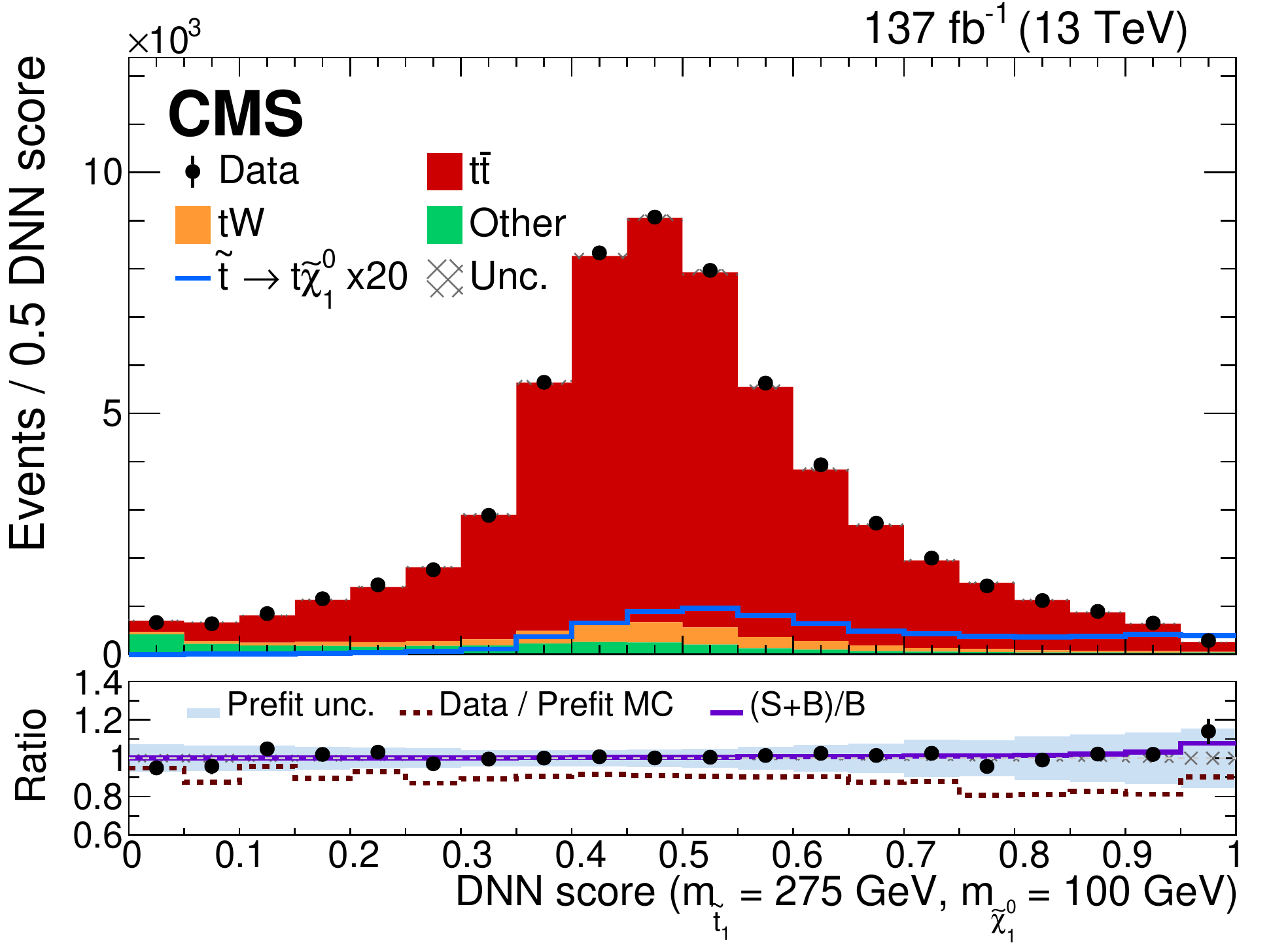}
\includegraphics[width=1\cmsFigWidth]{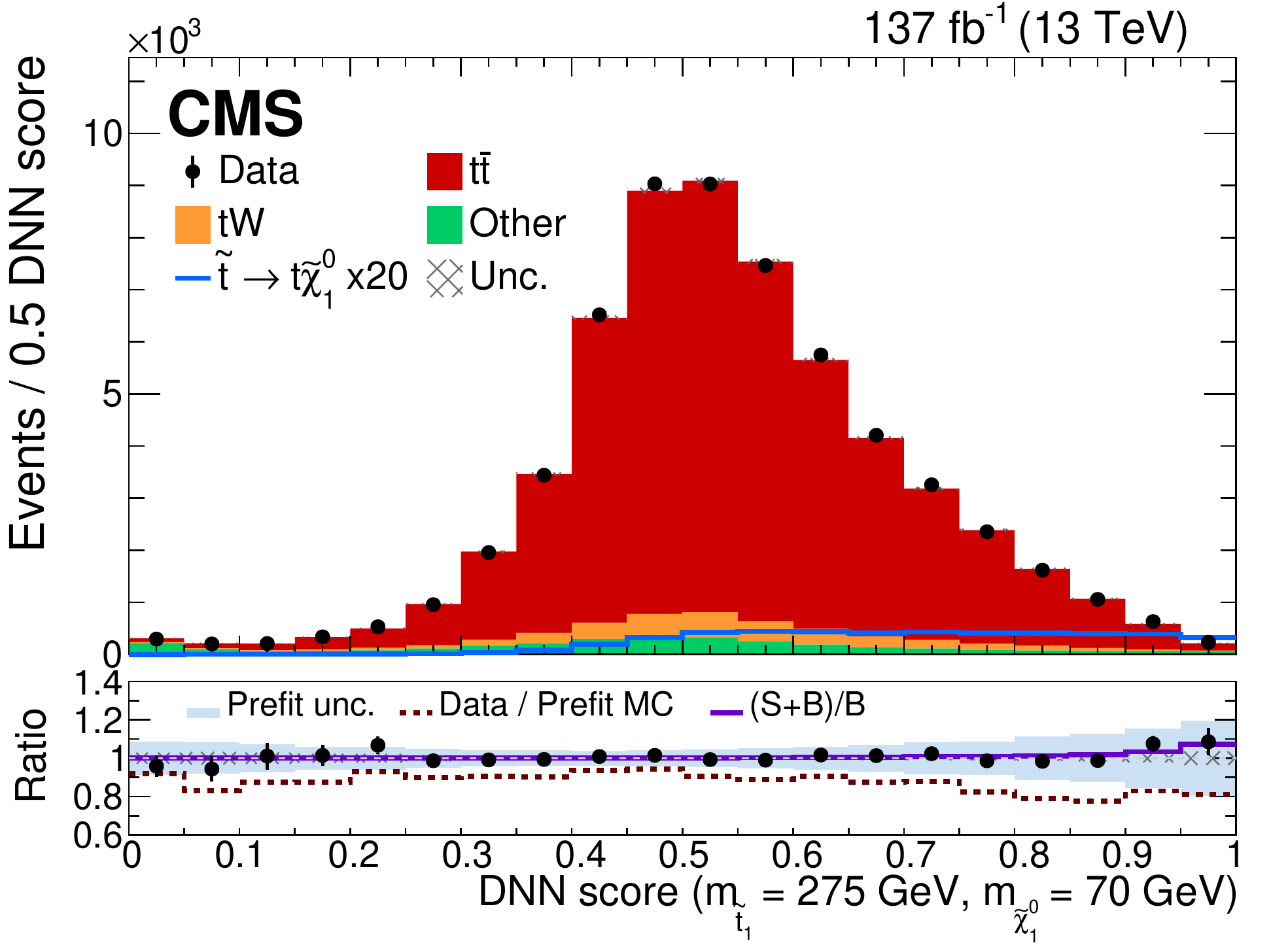}
\includegraphics[width=1\cmsFigWidth]{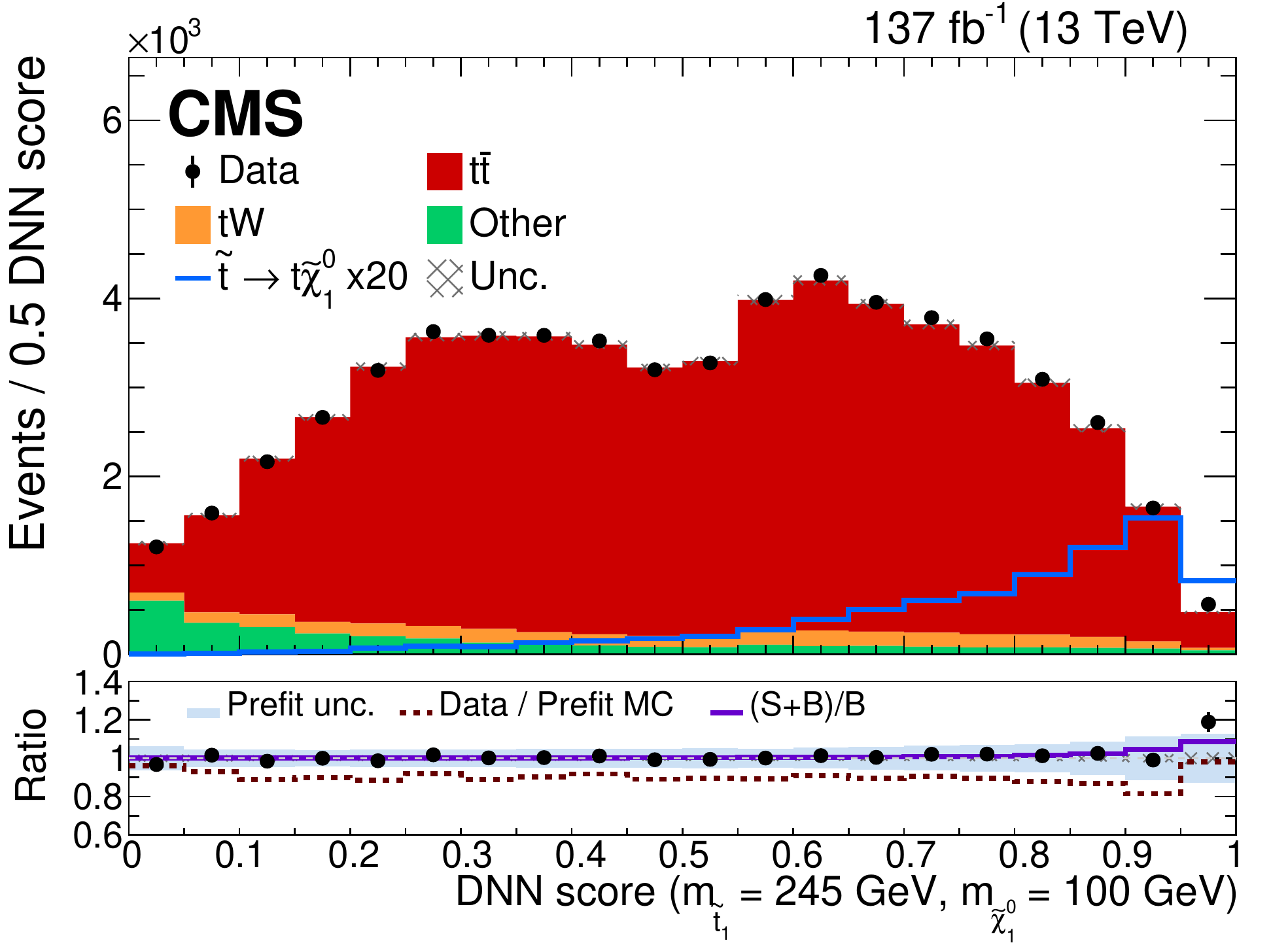}
  \caption{Post-fit DNN score distributions in the signal region for different mass hypotheses of, from upper left to lower right, $(\mstop, \mchi)=$ (225, 50); (275, 100); (275, 70); and (245, 100)\GeV. The superimposed signal prediction is scaled by the post-fit signal strength and, in the upper panels, it is also multiplied by a factor 20 for better visibility. The post-fit uncertainty band (crosses) includes statistical, background normalization, and all systematic uncertainties described in Section~\ref{sec:unc}. Events from $\ttbar\PW$, $\ttbar\PZ$, DY, nonprompt leptons, and diboson processes are grouped into the 'Other' category. The lower panel contains the data-to-prediction ratio before the fit (dotted brown line) and after (dots), each of them with their corresponding band of uncertainties (blue band for the pre-fit and crosses band for the post-fit). The ratio between the sum of the signal and background predictions and the background prediction (purple line) is also shown. The masses of the signal model correspond to the values of the DNN mass parameters in each distribution. }
\label{fig:dnnshape}
\end{figure*}

The statistical interpretation is performed by testing the SM hypothesis against the SUSY hypothesis. The data and predicted distributions for the DNN response in the signal region are combined in the nine channels (3 data-taking period x 3 lepton flavor combinations of the two leading leptons) in order to maximize the sensitivity to the signal. Each of the distributions is computed for different values of the mass parameters and compared to the prediction for the signal model with the corresponding masses. 
In Fig.~\ref{fig:dnnshape} the DNN score distributions for data are compared with those from the fit.  The fit function includes the background, and the signal prediction scaled by the post-fit signal strength, for different mass parameters. The points whose DNN distributions appear in the upper plots lie along the center line of the corridor, $\deltamcor = 0$, while those shown in the lower plots lie on its boundary.

A binned profile likelihood fit of the DNN output distribution is performed, where the nuisance parameters are modeled using Gaussian distributions. The correlation scheme for different data periods is described in Section~\ref{sec:unc}. No significant excess is observed over the background prediction for any of the distributions.

Upper limits on the production cross section of top squark pairs are calculated at 95\% confidence level (\CL) using a modified frequentist approach and the \CLs criterion, implemented through an asymptotic approximation~\cite{Cowan:2010js,Junk1999,Read:2002hq,CMS-NOTE-2011-005}. 

\begin{figure*}[htb!]
  \centering
     \includegraphics[width=1.5\cmsFigWidth]{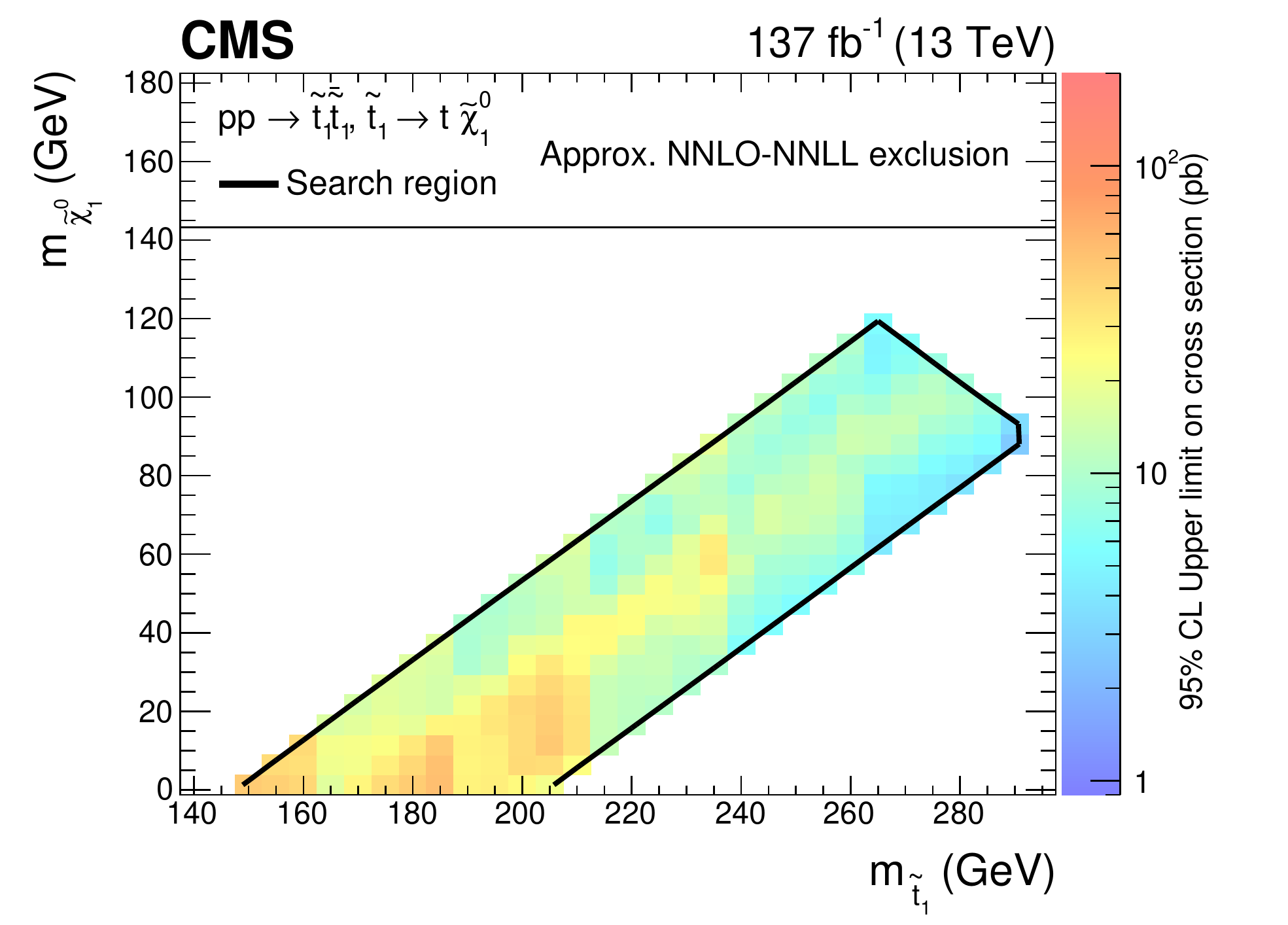}
     \caption{Upper limit at 95\% \CL on the signal cross section as a function of the top squark and neutralino masses in the top quark corridor region. The model is excluded for all of the colored region inside the black boundary.}
    \label{fig:rainbow}
\end{figure*}

\begin{figure*}[htb!]
  \centering
     \includegraphics[width=0.45\textwidth]{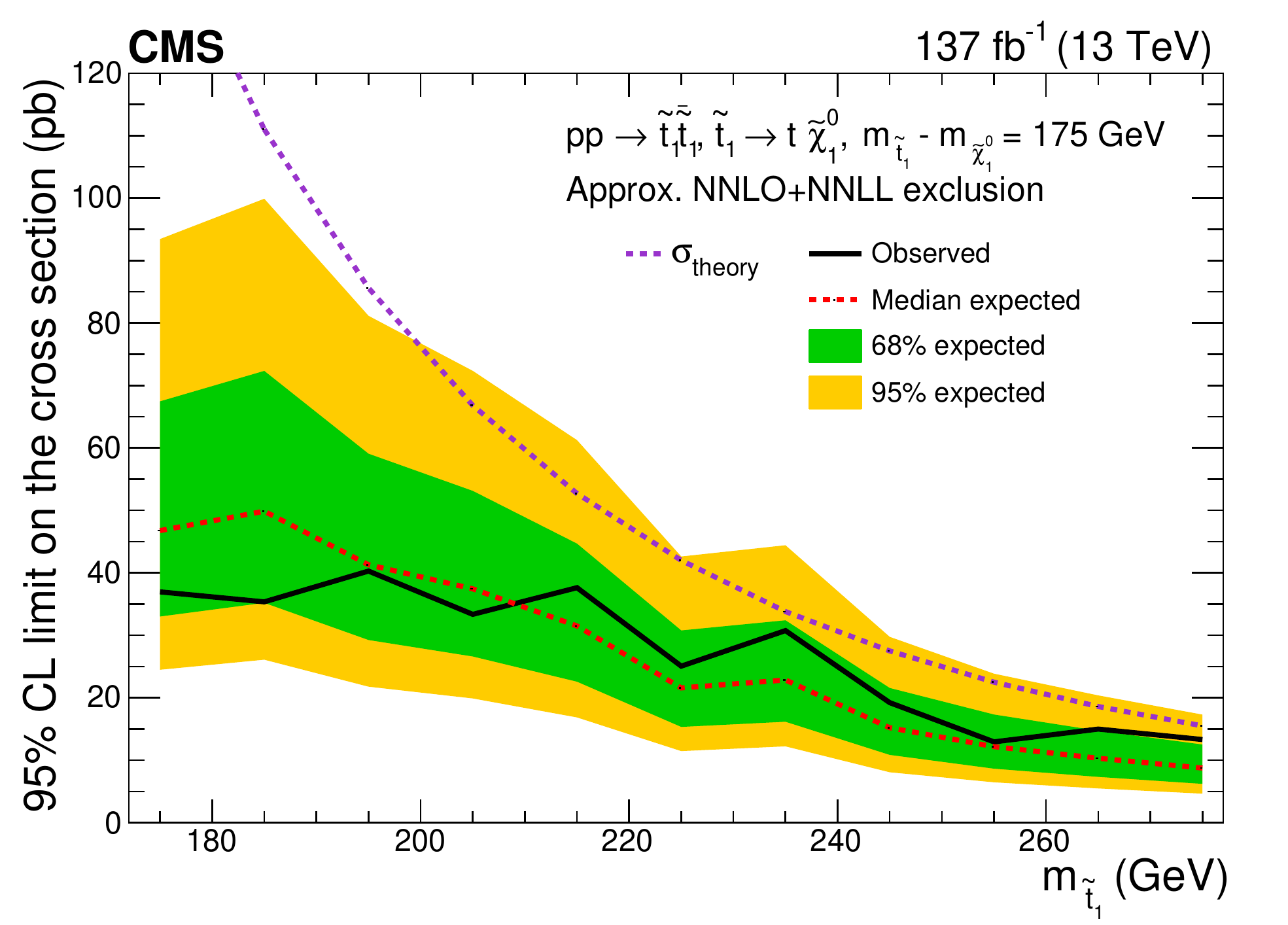}
     \includegraphics[width=0.45\textwidth]{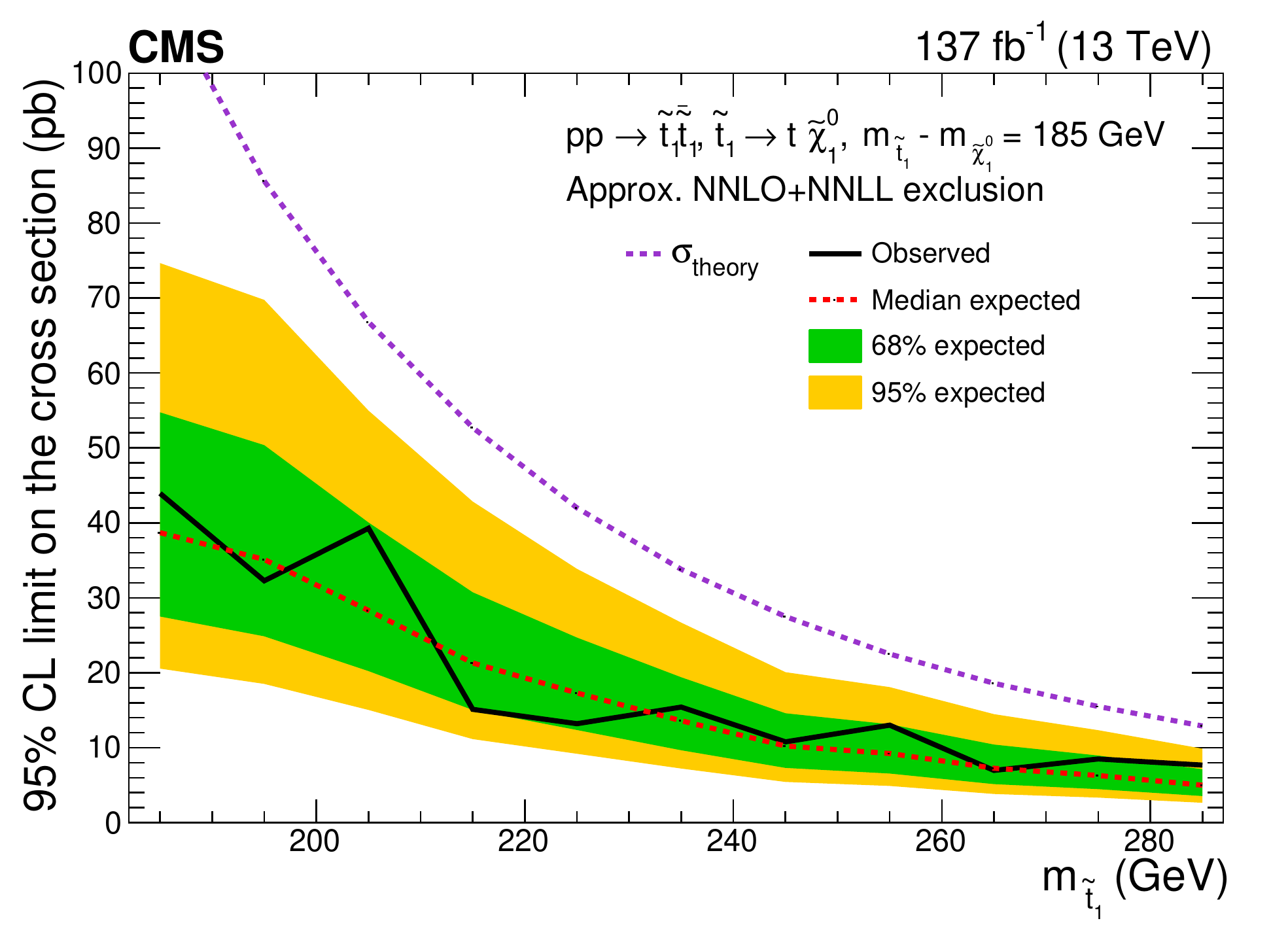}
     \includegraphics[width=0.45\textwidth]{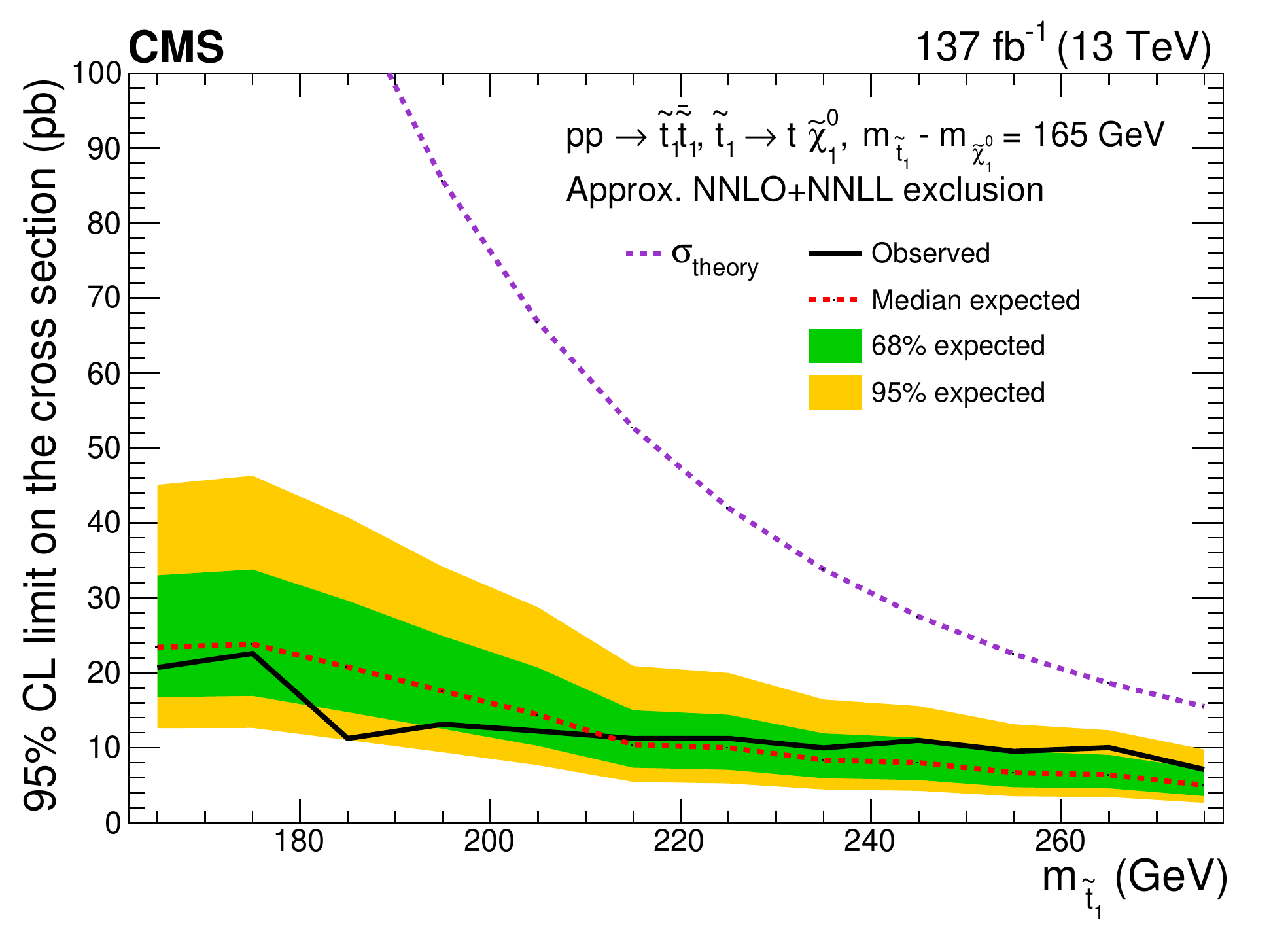}
     \caption{Upper limit at 95\% \CL on the signal cross section as a function of the top squark mass for \deltamsn of 175\GeV (upper left), 185\GeV (upper right) and 165\GeV (lower).
    The green and yellow bands represent the regions containing 68 and 95\%, respectively, of the distribution of limits expected under the background-only hypothesis.
    The purple dotted line indicates the approximate NNLO+NNLL production cross section. 
    }    
    \label{fig:brazilLimits}
\end{figure*}

Results are interpreted for different signals characterized by $ 145 < \mstop < 295\GeV$ and $\deltamcor\leq30\GeV$. The observed upper limit on the signal cross section is shown in Fig.~\ref{fig:rainbow}. The expected and observed upper limits are also shown for three different slices corresponding to $\deltamsn=165$, 175 and 185\GeV in Fig.~\ref{fig:brazilLimits}. Both the observed and expected cross section limits exclude the model over the region of the search.

\subsection{Combined results}

A statistical combination of the results of the three searches described in detail in Section~\ref{sec:bulk} is performed outside the corridor area in order to provide interpretations in the context of the signal scenarios described in Section~\ref{intro}.
The signal regions of the analyses targeting different final states are designed to be mutually exclusive.
Additionally, there is no significant overlap of any of the control regions with signal regions of a different analysis.
The overlap between control regions of the single-lepton and dilepton analysis is found to be less than 1\% and therefore considered negligible.
Correlations of systematic uncertainties in the expected signal and background yields are studied and taken into account.
Uncertainties in the jet energy scale and \ptmiss resolution, \PQb~tagging efficiency scale factors, heavy resonance taggers, integrated luminosity and background normalizations are treated as fully correlated.
Because of differences in the lepton identification methods and working points, as well as the triggers to select events, the corresponding uncertainties are considered uncorrelated.
Theory uncertainties in the choice of the PDF, ${\mu}_\text{R}$ and ${\mu}_\text{F}$ and ISR modeling of the signal prediction, as well as SM backgrounds that are estimated using simulation, are taken to be fully correlated.

\begin{figure*}[htb!]
  \centering
     \includegraphics[width=0.45\textwidth]{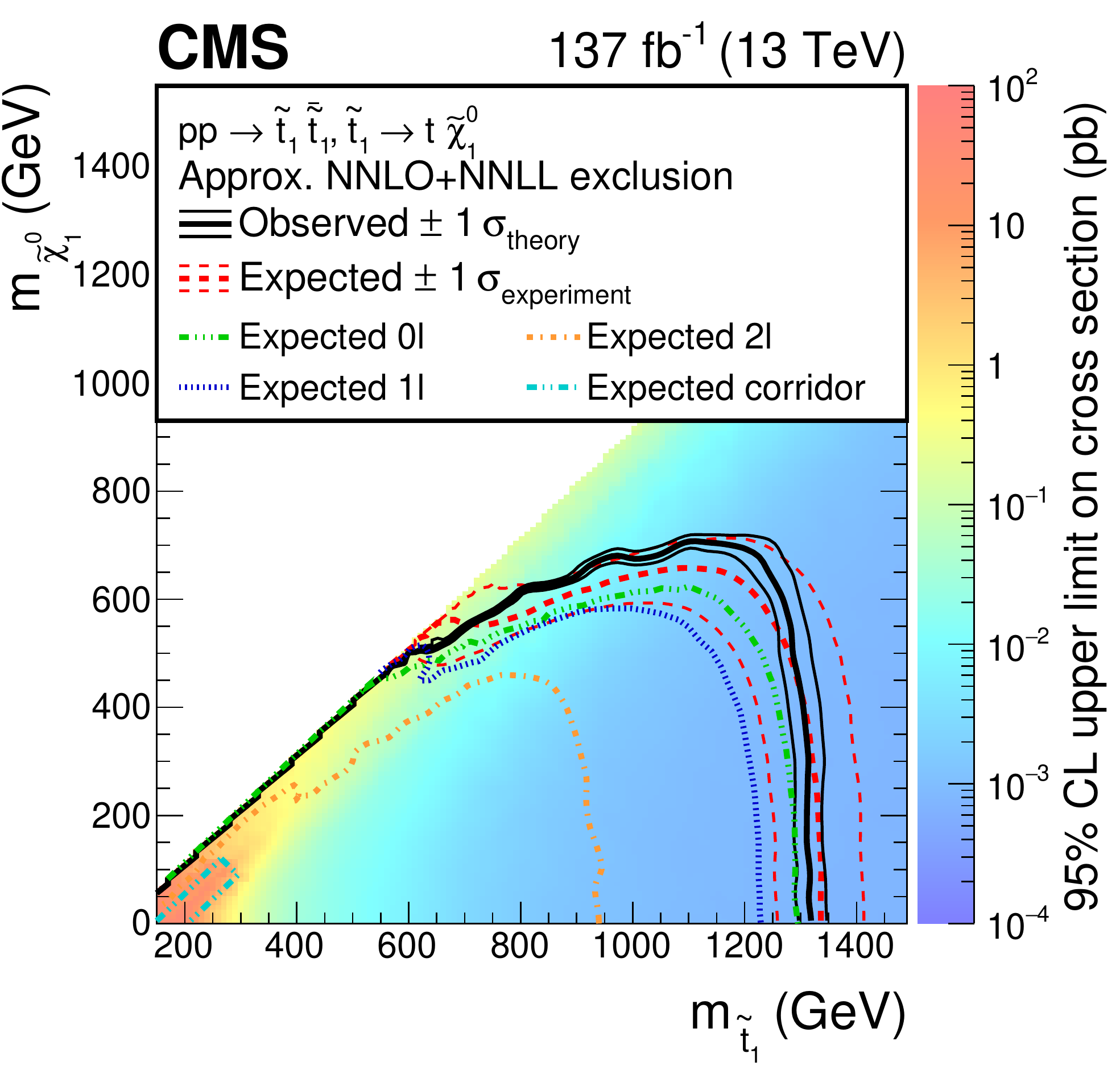}
     \includegraphics[width=0.45\textwidth]{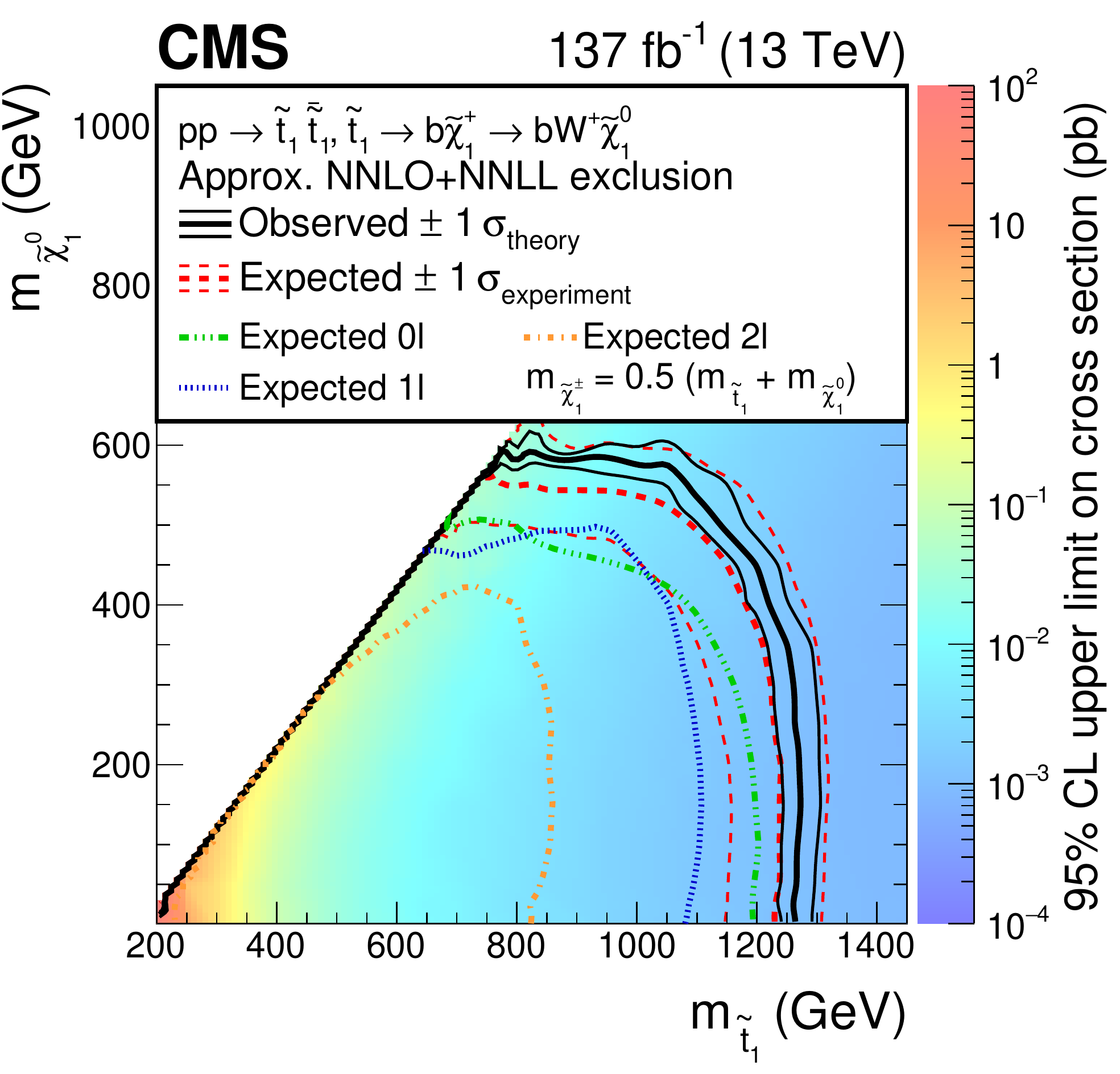}
     \includegraphics[width=0.45\textwidth]{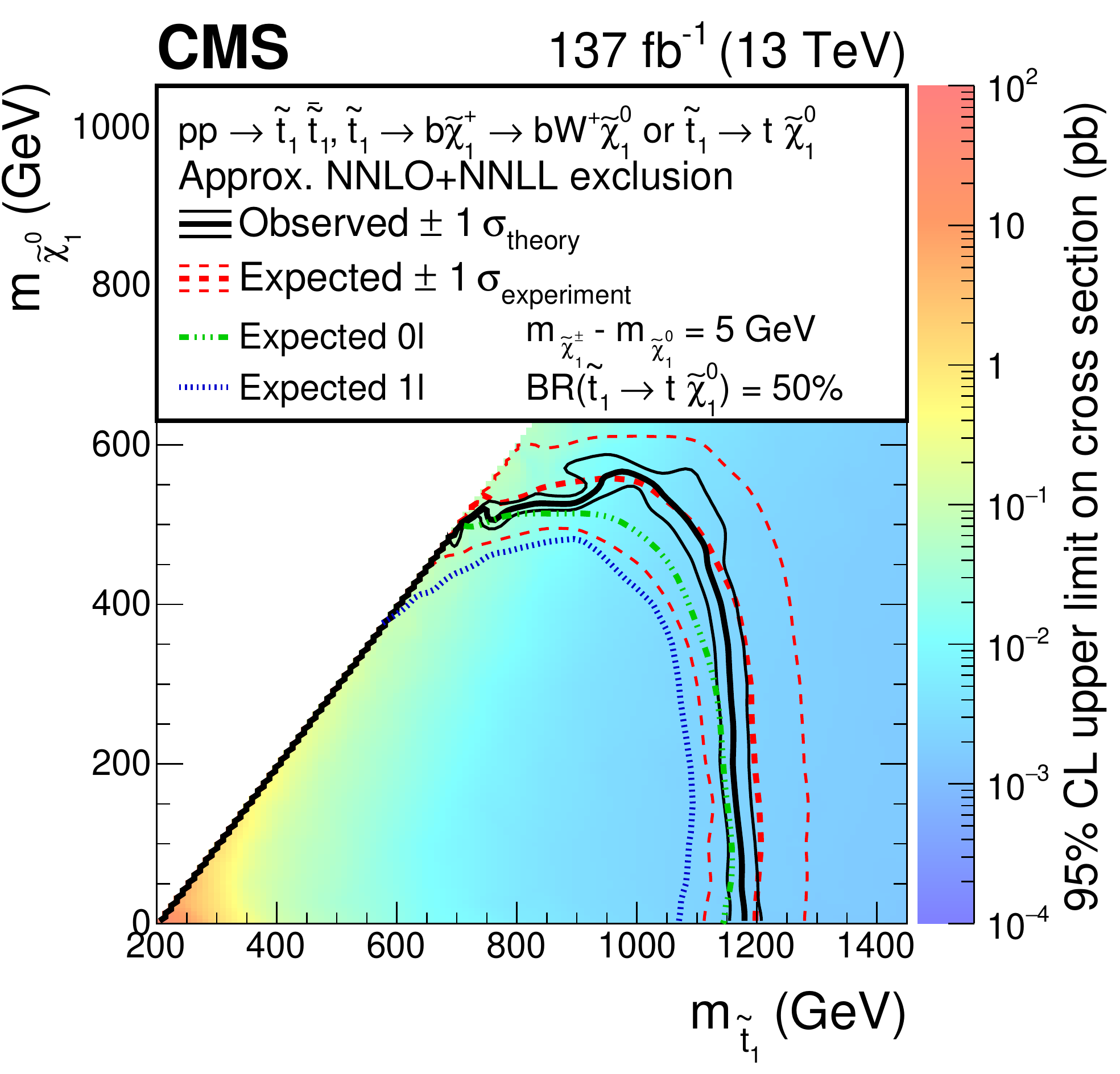}
     \caption{
     Expected and observed limits in the $m_{\PSQtDo}$-$m_{\PSGczDo}$ mass plane, for the $\PSQtDo \to \PQt \PSGczDo$ model (upper left), the $\PSQtDo \to \PQb\PSGcpDo \to \PQb\PWp\PSGczDo$ model (upper right) and a model with a branching fraction of 50\% for each of these top squark decay modes (lower), assuming a mass difference between the neutralino and chargino of 5\GeV.
     The color indicates the 95\% \CL upper limit on the cross section at each point in the plane.
     The area below the thick black curve represents the observed exclusion region at 95\% \CL,
     while the dashed red lines indicate the expected limits at 95\% \CL and the region containing 68\% of the distribution of limits expected under the background-only hypothesis of the combined analyses.
     The thin black lines show the effect of the theoretical uncertainties in the signal cross section.
     }
    \label{fig:T2tt_limit}
\end{figure*}

Figure~\ref{fig:T2tt_limit} (upper left) shows the combination of the results of the three searches for direct top squark pair production for the model with $\PSQtDo \to \PQt \PSGczDo$ decays.
The analysis described in Section~\ref{sec:corridor} is exclusively used for extracting limits in the top quark corridor region.
No result of the other analyses is used in this particular region of parameter space.
The combined result excludes a top squark mass of 1325\GeV for a massless LSP, and an LSP mass of 700\GeV for a top squark mass of 1150\GeV.
The expected limit of the combination is dominated by the fully hadronic search for signals with large mass splitting.
In regions with smaller mass splitting between the top squark and the LSP, searches in the zero- and single-lepton channels have similar sensitivity.

Figure~\ref{fig:T2tt_limit} (upper right) shows the equivalent limits for direct top squark pair production for the model with  $\PSQtDo \to \PQb\PSGcpDo \to \PQb\PWp\PSGczDo$ decays.
The mass of the chargino is set to the mean of the masses of the top squark and the LSP.
The combined result for this scenario excludes a top squark mass of 1260\GeV for a massless LSP and an LSP mass of 575\GeV for a top squark mass of 1000\GeV.
The combination extends the sensitivity to both top squark and LSP masses by about 50\GeV compared to the most sensitive individual result coming from the fully hadronic channel.

Figure~\ref{fig:T2tt_limit} (lower) shows the limits for the model with a 50\% branching fraction of the top squark decays discussed previously.
In this model, the mass splitting between the neutralino and chargino is assumed to be 5\GeV.
Because of the low acceptance for low-momentum leptons the dilepton result is not interpreted in terms of this model.
Top squark masses up to 1175\GeV are excluded in this model when the LSP is massless, and up to 1000\GeV for LSP masses up to 570\GeV.

As shown in Fig.~\ref{fig:T2tt_limit} (upper left), the region of the parameter space of the simplified SUSY models considered for interpretation in this analysis, which is favored by the naturalness paradigm, is now further constrained by the exclusion limits.

\subsection{Search for dark matter in association with top quarks}
\begin{figure*}[htb!]
  \centering
     \includegraphics[width=0.45\textwidth]{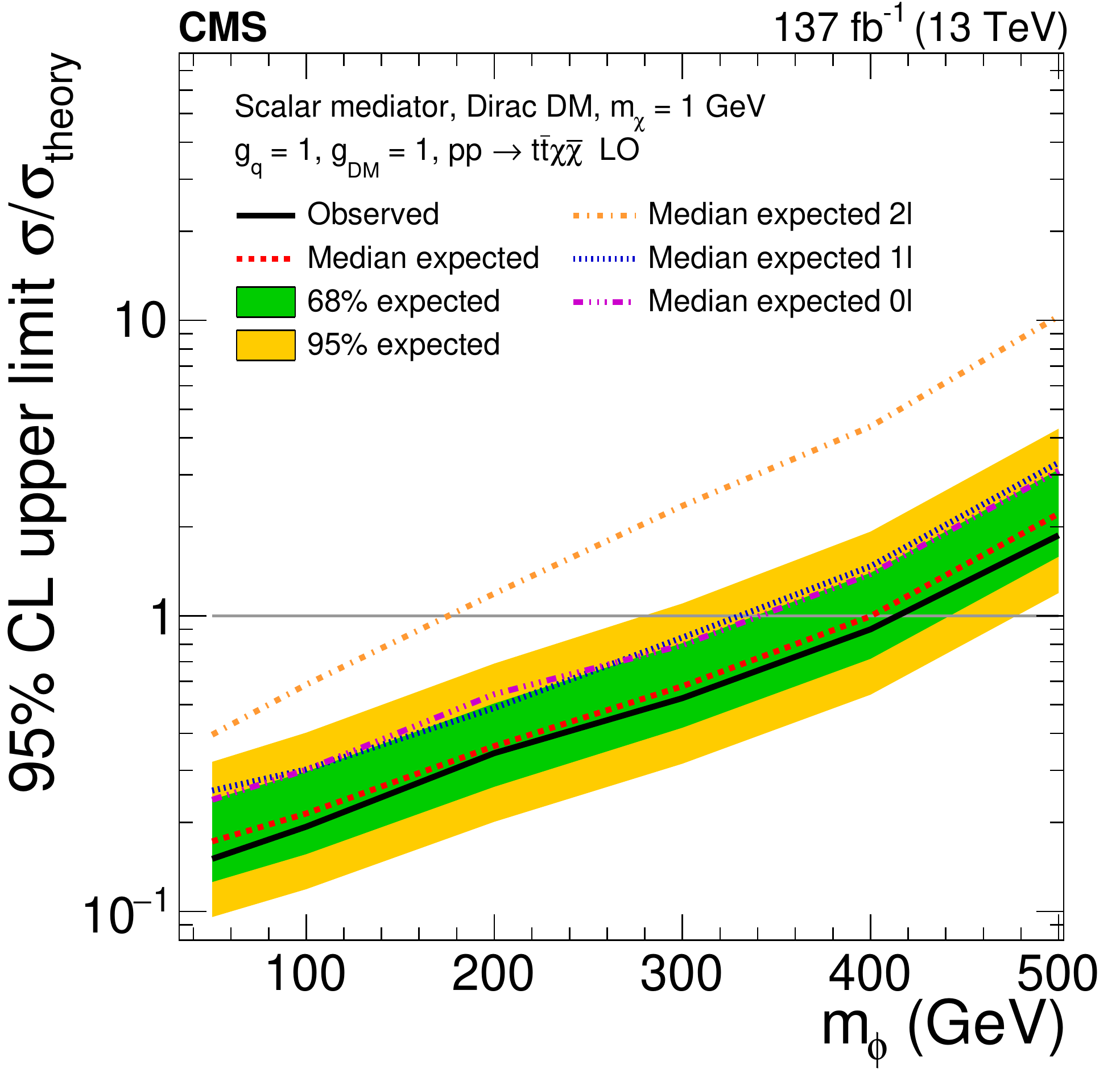}
     \includegraphics[width=0.45\textwidth]{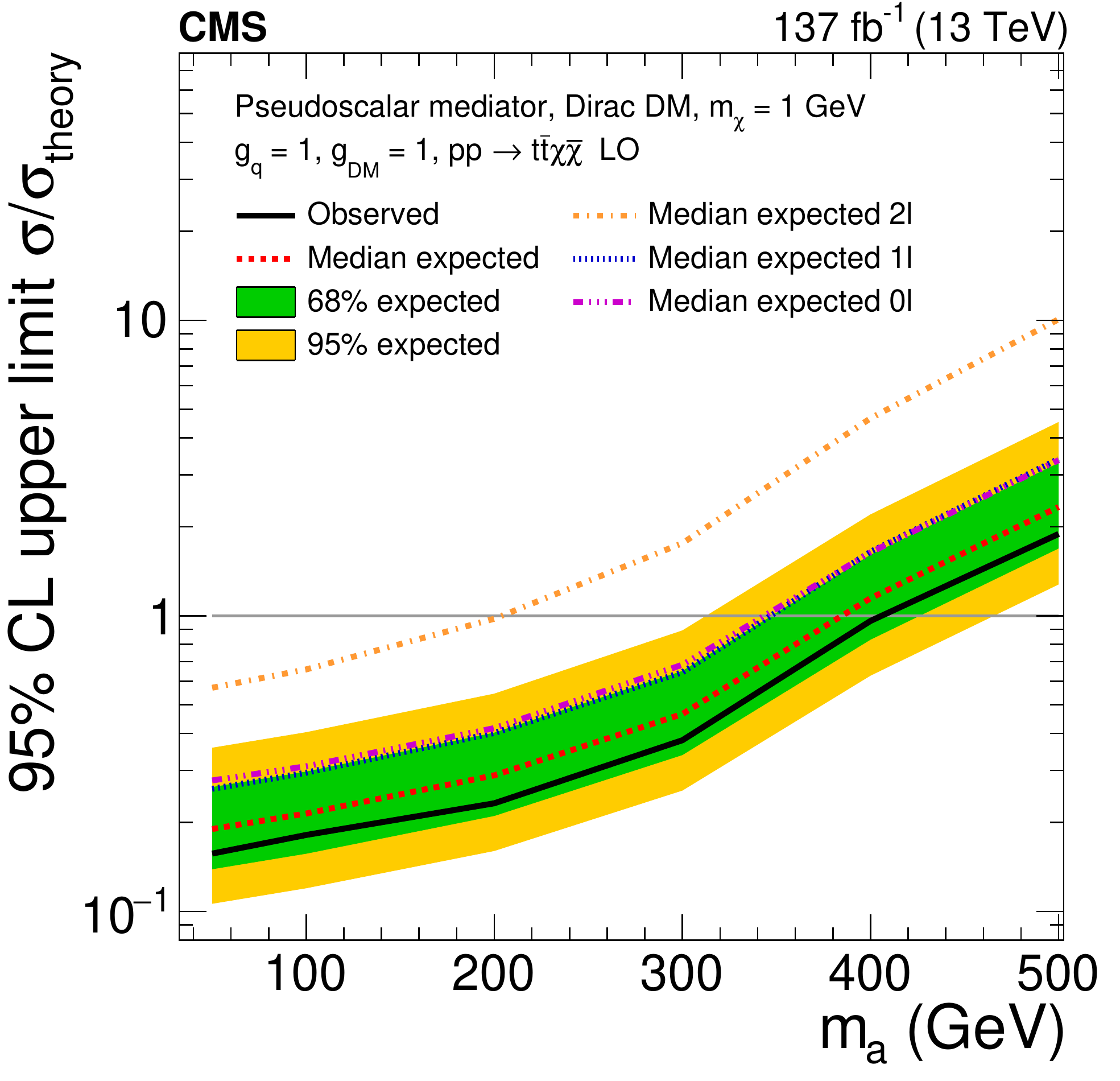}
     \caption{The 95\% \CL expected (dashed line) and observed limits (solid line) on $\sigma/\sigma_{\mathrm{theory}}$ for a fermionic DM particle with $m_{\chi}=1\GeV$, as a function of the mediator mass for a scalar (left) and pseudoscalar (right).
    The green and yellow bands represent the regions containing 68 and 95\%, respectively, of the distribution of limits expected under the background-only hypothesis.
    The horizontal gray line indicates $\sigma/\sigma_{\mathrm{theory}}=1$.
    The mediator couplings are set to $g_\Pq=g_{\mathrm{DM}}=1$.
     }
    \label{fig:ttDM_limit}
\end{figure*}

The results of the inclusive top squark searches are interpreted in simplified models of associated production of DM particles with a top quark pair, shown in Fig.~\ref{fig:TTDMdiagrams}.
The interaction of the DM particles and the top quark is mediated by a scalar or pseudoscalar mediator particle.
Assuming a dark matter particle mass of 1\GeV, scalar and pseudoscalar mediators with masses up to 400 and 420\GeV are excluded at 95\% CL, respectively, as shown in Fig.~\ref{fig:ttDM_limit}.
The obtained upper limits on $\sigma(\Pp\Pp\to\ttbar\chi\tilde{\chi})/\sigma_{\mathrm{theory}}$ are independent of the mass of the DM fermion ($m_{\chi}$), as long as the mediator is produced on-shell~\cite{Abercrombie:2015wmb}.
Previous results are improved by more than 100\GeV~\cite{Sirunyan:2018dub,Sirunyan:2019gfm} and the sensitivity extends beyond $m_{\phi/a}>2 m_{\PQt}$ for the first time.
The competing decay channel of the mediator into a top quark pair, $\phi/a \to \ttbar$, is taken into account in the signal simulation and cross section calculation.

\section{Summary}

Four searches for top squark pair production and their statistical combination are presented.
The searches use a data set of proton-proton collisions at a center-of-mass energy of 13\TeV collected by the CMS detector and corresponding to an integrated luminosity of 137\fbinv.
A dedicated analysis is presented that is sensitive to signal models where the mass splitting between the top squark and the lightest supersymmetric particle (LSP) is close to the top quark mass.
A deep neural network algorithm is used to separate the signal from the top quark background using events containing an opposite-charge dilepton pair, at least two jets, at least one \cPqb-tagged jet, $\ptmiss>50\GeV$, and stransverse mass greater than 80\GeV.
No excess of data over the standard model prediction is observed, and upper limits are set at 95\% confidence level on the top squark production cross section.
Top squarks with mass from 145 to 275\GeV, for LSP mass from 0 to 100\GeV, with a mass difference between the top squarks and LSP of up to 30\GeV deviation around the mass of the top quark, are excluded for the first time in CMS.
Previously published searches in final states with 0, 1, or 2 leptons are combined to extend the exclusion limits of top squarks with masses up to 1325\GeV for a massless LSP and an LSP mass up to 700\GeV for a top squark mass of 1150\GeV, for certain models of top squark production.
In an alternative signal model of dark matter production via a spin-0 mediator in association with a top quark pair, mediator particle masses up to 400 and 420\GeV are excluded for scalar or pseudoscalar mediators, respectively, assuming a dark matter particle mass of 1\GeV.

\begin{acknowledgments}

  We congratulate our colleagues in the CERN accelerator departments for the excellent performance of the LHC and thank the technical and administrative staffs at CERN and at other CMS institutes for their contributions to the success of the CMS effort. In addition, we gratefully acknowledge the computing  centers and personnel of the Worldwide LHC Computing Grid and other  centers for delivering so effectively the computing infrastructure essential to our analyses. Finally, we acknowledge the enduring support for the construction and operation of the LHC, the CMS detector, and the supporting computing infrastructure provided by the following funding agencies: BMBWF and FWF (Austria); FNRS and FWO (Belgium); CNPq, CAPES, FAPERJ, FAPERGS, and FAPESP (Brazil); MES (Bulgaria); CERN; CAS, MoST, and NSFC (China); MINCIENCIAS (Colombia); MSES and CSF (Croatia); RIF (Cyprus); SENESCYT (Ecuador); MoER, ERC PUT and ERDF (Estonia); Academy of Finland, MEC, and HIP (Finland); CEA and CNRS/IN2P3 (France); BMBF, DFG, and HGF (Germany); GSRT (Greece); NKFIA (Hungary); DAE and DST (India); IPM (Iran); SFI (Ireland); INFN (Italy); MSIP and NRF (Republic of Korea); MES (Latvia); LAS (Lithuania); MOE and UM (Malaysia); BUAP, CINVESTAV, CONACYT, LNS, SEP, and UASLP-FAI (Mexico); MOS (Montenegro); MBIE (New Zealand); PAEC (Pakistan); MSHE and NSC (Poland); FCT (Portugal); JINR (Dubna); MON, RosAtom, RAS, RFBR, and NRC KI (Russia); MESTD (Serbia); SEIDI, CPAN, PCTI, and FEDER (Spain); MOSTR (Sri Lanka); Swiss Funding Agencies (Switzerland); MST (Taipei); ThEPCenter, IPST, STAR, and NSTDA (Thailand); TUBITAK and TAEK (Turkey); NASU (Ukraine); STFC (United Kingdom); DOE and NSF (USA).
  
  \hyphenation{Rachada-pisek} Individuals have received support from the Marie-Curie program and the European Research Council and Horizon 2020 Grant, contract Nos.\ 675440, 724704, 752730, 758316, 765710, 824093, and COST Action  CA16108 (European Union); the Leventis Foundation; the Alfred P.\ Sloan Foundation; the Alexander von Humboldt Foundation; the Belgian Federal Science Policy Office; the Fonds pour la Formation \`a la Recherche dans l'Industrie et dans l'Agriculture (FRIA-Belgium); the Agentschap voor Innovatie door Wetenschap en Technologie (IWT-Belgium); the F.R.S.-FNRS and FWO (Belgium) under the ``Excellence of Science -- EOS" -- be.h project n.\ 30820817; the Beijing Municipal Science \& Technology Commission, No. Z191100007219010; the Ministry of Education, Youth and Sports (MEYS) of the Czech Republic; the Deutsche Forschungsgemeinschaft (DFG), under Germany's Excellence Strategy -- EXC 2121 ``Quantum Universe" -- 390833306, and under project number 400140256 - GRK2497; the Lend\"ulet (``Momentum") Program and the J\'anos Bolyai Research Scholarship of the Hungarian Academy of Sciences, the New National Excellence Program \'UNKP, the NKFIA research grants 123842, 123959, 124845, 124850, 125105, 128713, 128786, and 129058 (Hungary); the Council of Science and Industrial Research, India; the Latvian Council of Science; the Ministry of Science and Higher Education and the National Science Center, contracts Opus 2014/15/B/ST2/03998 and 2015/19/B/ST2/02861 (Poland); the National Priorities Research Program by Qatar National Research Fund; the Ministry of Science and Higher Education, project no. 0723-2020-0041 (Russia); the Programa Estatal de Fomento de la Investigaci{\'o}n Cient{\'i}fica y T{\'e}cnica de Excelencia Mar\'{\i}a de Maeztu, grant MDM-2015-0509 and the Programa Severo Ochoa del Principado de Asturias; the Stavros Niarchos Foundation (Greece); the Rachadapisek Sompot Fund for Postdoctoral Fellowship, Chulalongkorn University and the Chulalongkorn Academic into Its 2nd Century Project Advancement Project (Thailand); the Kavli Foundation; the Nvidia Corporation; the SuperMicro Corporation; the Welch Foundation, contract C-1845; and the Weston Havens Foundation (USA).
\end{acknowledgments}

\bibliography{auto_generated}
\cleardoublepage \appendix\section{The CMS Collaboration \label{app:collab}}\begin{sloppypar}\hyphenpenalty=5000\widowpenalty=500\clubpenalty=5000\vskip\cmsinstskip
\textbf{Yerevan Physics Institute, Yerevan, Armenia}\\*[0pt]
A.~Tumasyan
\vskip\cmsinstskip
\textbf{Institut f\"{u}r Hochenergiephysik, Wien, Austria}\\*[0pt]
W.~Adam, J.W.~Andrejkovic, T.~Bergauer, S.~Chatterjee, M.~Dragicevic, A.~Escalante~Del~Valle, R.~Fr\"{u}hwirth\cmsAuthorMark{1}, M.~Jeitler\cmsAuthorMark{1}, N.~Krammer, L.~Lechner, D.~Liko, I.~Mikulec, P.~Paulitsch, F.M.~Pitters, J.~Schieck\cmsAuthorMark{1}, R.~Sch\"{o}fbeck, M.~Spanring, S.~Templ, W.~Waltenberger, C.-E.~Wulz\cmsAuthorMark{1}
% ...
\vskip\cmsinstskip
% ...
1:  Also at TU Wien, Wien, Austria\\
% ...
\end{sloppypar}
\end{document}